\documentclass[twocolumn,amsmath,amssymb,floatfix,superscriptaddress,prb]{revtex4-2}
\usepackage{url,epsf}
\usepackage{comment}
\usepackage{appendix}
\usepackage{physics}
\usepackage{multirow}
\usepackage[colorlinks=true,linkcolor=blue]{hyperref}
\usepackage{amsmath,bm,graphicx,epsfig}
\usepackage{xcolor}
\usepackage[english]{babel}
\usepackage{amsfonts,stmaryrd,amssymb}
\usepackage{enumerate}
\usepackage{csquotes}
\usepackage{ulem}

\begin{document}

\title{
Topological Kondo semimetal and insulator in AB-stacked heterobilayer transition metal dichalcogenides
}

\author{Daniele Guerci}
\affiliation{Center for Computational Quantum Physics, Flatiron Institute, New York, New York 10010, USA}

\author{Kevin P. Lucht}
\affiliation{Department of Physics and Astronomy, Center for Materials Theory, Rutgers University, Piscataway, New Jersey 08854, USA}

\author{Valentin Cr\'epel}
\affiliation{Center for Computational Quantum Physics, Flatiron Institute, New York, New York 10010, USA}

\author{Jennifer Cano}
\affiliation{Department of Physics and Astronomy, Stony Brook University, Stony Brook, New York 11794, USA}
\affiliation{Center for Computational Quantum Physics, Flatiron Institute, New York, New York 10010, USA}

\author{J. H. Pixley}
\affiliation{Department of Physics and Astronomy, Center for Materials Theory, Rutgers University, Piscataway, New Jersey 08854, USA}
\affiliation{Center for Computational Quantum Physics, Flatiron Institute, New York, New York 10010, USA}

\author{Andrew Millis}
\affiliation{Center for Computational Quantum Physics, Flatiron Institute, New York, New York 10010, USA}
\affiliation{Department of Physics, Columbia University, 538 West 120th Street, New York, NY 10027}

\begin{abstract}

   Recent experiments reported the realization of a heavy Fermi liquid in AB-stacked MoTe$_2$/WSe$_2$ heterobilayers.
   In this paper we show that the AB-stacked heterobilayer configuration is particularly suited to realize topological Kondo semimetal and topological Kondo insulator ground states at a doping of two holes per moir\'e unit cell. The small lattice mismatch between the MoTe$_2$ and WSe$_2$ monolayers and the different bandwidths of their highest lying  moir\'e valence bands  means that, in the experimentally relevant range of hole dopings, the MoTe$_2$ layer is effectively a Mott insulator with only low-lying magnetic excitations Kondo-coupled to more itinerant electrons in the WSe$_2$. The crucial consequence of the AB-stacking configuration is that the interlayer tunnelling connects orbitals of opposite parity in the two layers, leading to a chiral Kondo coupling. 
  We show that the chiral Kondo coupling favors a topological Kondo semimetal at filling $\nu=1+1$, with a non-quantized spin Hall conductance arising from edge modes, whose spectrum and overlap with bulk states we determine. We further show that a spatially random strain field that locally breaks the rotation symmetry can convert the Kondo semimetal to a narrow gap topological Kondo insulator featuring a quantized spin Hall conductance. 

\end{abstract}

\maketitle

\section{Introduction} 


Transition metal dichalcogenides (TMDs) are wide-gap semiconductors displaying a strong Ising spin-orbit coupling (SOC)  in the valence band~\cite{kormanyos2015k} with (in most cases) a honeycomb lattice structure and valence band maxima located at the corners of the hexagonal Brillouin zone ($K$ and $K^\prime$ points in the usual hexagonal notation).   In this circumstance, the SOC leads to an effective spin-valley locking: carriers at the top of the valence band in valley $K$/$K'$ have spin-$\downarrow$/$\uparrow$ and carry spin and orbital angular momentum $L^z$ reflecting the character $d_{\pm}=d_{x^2-y^2}\pm i d_{xy}$ of each monolayer~\cite{kormanyos2015k}. Because lightly $p$-doped TMD monolayers behave as a two-dimensional two component Fermi gas, where the two fermionic flavours correspond to states at the $K$ and $K'$ points of the monolayer Brillouin zone, TMDs have become central elements in the construction of two-dimensional moir\'e materials that host  a wide variety of emergent strongly correlated phenomena including  Mott transitions~\cite{Wang_2020,Li_2021}, superconductivity~\cite{xia2024unconventionalsuperconductivitytwistedbilayer,guo2024superconductivitytwistedbilayerwse2}, quantum criticality~\cite{Ghiotto_2021} and integer and fractional anomalous Hall phases~\cite{Cai_2023,Park_2023,zeng2023thermodynamic,XuFan_2023,Foutty_2024,kang2024observationdoublequantumspin,kang2024observationfractionalquantumspin}.

A particularly interesting class of moir\'e compound are the AB-stacked heterobilayer transition metal dichalcogenides (TMDs), obtained by stacking two different TMDs at a $60^\circ$ twist rotation~\cite{mak2022semiconductor,zhao2022realizationhaldanecherninsulator,tao2022valleycoherentquantumanomaloushall,tao2023giantspinhalleffect}. The small lattice mismatch between the two components leads to a long-period moir\'e structure while the $60^\circ$ twist means the two valence bands coupled by the moir\'e potential  are characterized by opposite $L^z$~\cite{Zhang_2021} so that the interlayer tunneling vanishes at the $K/K^\prime$ points and acquires a winding from the angular momentum mismatch. The winding gives rise to non-trivial topological properties~\cite{zhao2022realizationhaldanecherninsulator,tao2023giantspinhalleffect}. Physical differences between the two component materials constituting the heterobilayer lead to an energy offset that is typically large enough for the first holes injected into the heterostructure to primarily go into one layer. In the MoTe$_2$/WSe$_2$  case of interest here, the holes first populate the MoTe$_2$ layer and correlations in this layer are strong enough that at a hole density of $\nu=1$ per  moir\'e unit cell  the state is a correlation driven (Mott) insulator~\cite{Li_2021}. However the energy offset may be experimentally tuned by applying a perpendicular ``displacement" electric field leading at $\nu=1$ to a crossover from a Mott insulator (gap controlled by the in-layer Coulomb interaction) to a charge transfer insulator (lowest hole addition state is on the other layer) ~\cite{Devakul_2021,Guerci_2023}. Tuning yet further,  the system realizes a transition from a layer polarized Mott insulator to a quantum anomalous Hall insulator~\cite{Devakul_2021,Zhang_2021,Pan_2022,Xie_2022,Chang_2022,Dong_2023,Xie_2023}. 

In the charge transfer insulator regime, increasing the hole density beyond $\nu=1$ to $\nu=\nu_{\rm Mo}+\nu_{\rm W}=1+x$ with $x>0$ gives rise to a ``heavy fermion" situation in which a layer of local moments in the MoTe$_2$ layer coexists with itinerant carriers in the WSe$_2$ layer~\cite{Zhao_2023,zhao2023emergence,Dalal_2021,Kumar_2022,Guerci_2023,Xie_2024,GangPRR_2023}. As pointed out in our recent work~\cite{Guerci_2023}, the chiral nature of the interlayer exchange gives rise to a heavy Fermi liquid phase whose Fermi surface is pierced by a topological Berry flux. Moreover, at low-doping the  chiral exchange can stabilize a $\mathbb Z_2$ topological superconductor~\cite{Crepel_2023}. 

In this work we consider the state emerging when the hole concentration is increased to $\nu=2$ with $\nu_{\rm Mo}=\nu_{\rm W}=1$. We show that in the ideal case the system provides a realization of the topological Kondo semimetal phase~\cite{Pixley-2017,Chang-2018,lai2018weyl,Chen_2024} previously proposed by one of us and by others on theoretical grounds. 
The Kondo semimetal phase involves local moments with short range antiferromagnetic correlations in MoTe$_2$ exchange-coupled to itinerant carriers in WSe$_2$ leading to a spin Hall phase of heavy quasiparticles~\cite{Dzero_2010,Dzero_prb_2012,Dzero_2016}. We characterize the properties of the edge modes of the quantum spin Hall phase. 
We determine the range of stability of the Kondo semimetal phase in temperature and under various perturbations. Importantly, we show how random strain fields arising from interfacial disorder fill in the $p$-wave hybridization gap to realize a narrow gap topological Kondo insulator in a moir\'e material.  

The plan of the paper is as follows. In Sec.~\ref{sec:continuum_modeling} we present the Hamiltonian, highlighting the essential properties including the opposite parity of the Wannier orbitals and the spin-momentum locking that are key to realize the topological Kondo phase. In Sec.~\ref{sec:chiralPAM}, we present the strong-coupling limit of the lattice Hamiltonian and we introduce the slave-boson (`parton') theory that describes the physics in this limit. Sec.~\ref{sec:meanfieldtheory} is devoted to the mean-field solution of the model where we detail the criterion for the formation of the heavy Fermi liquid phase. In Sec.~\ref{sec:semimetal} we characterize the topological Kondo semimetal phase, while in Sec.~\ref{sec:strain} we show how uniform or random strain fields may open a full gap resulting in a quantum spin Hall Kondo insulator. 
In Sect.~\ref{sec:edgemodes} we examine the edge modes of the Kondo semimetal quantifying their overlap with bulk states in the topological Kondo semi-metallic regime by studying a system with cylindrical boundary conditions open in one direction and periodic in the other. Finally, we provide a summary of the main results of our work and outlook in Sec.~\ref{sec:conclusions}. Appendices provide details of some of the calculations in the main text.

\begin{figure}
    \centering
    \includegraphics[width=0.8\linewidth]{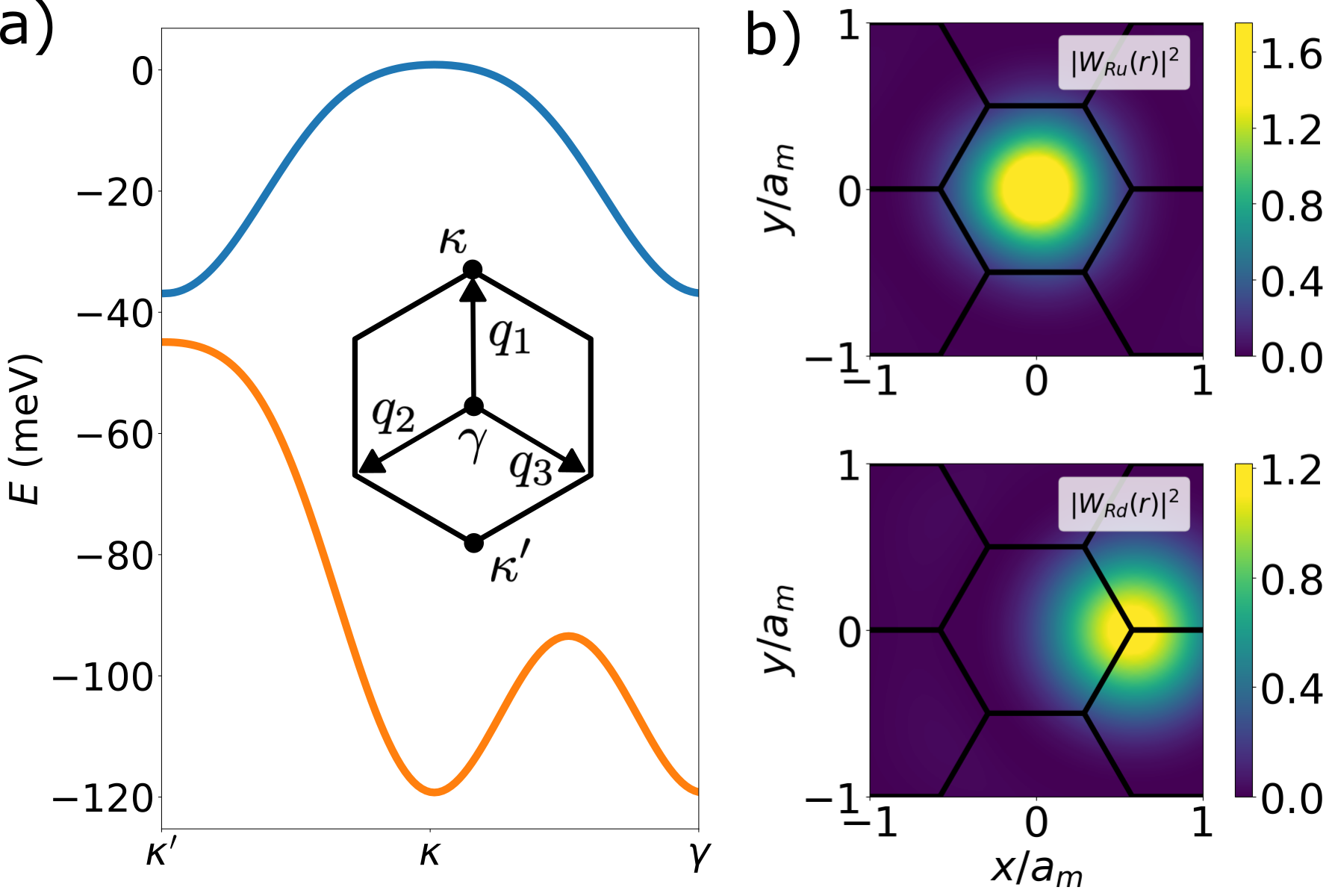}
    \vspace{-0.2cm}
    \caption{a) Tight binding model band structure of AB-stacked MoTe$_2$/WSe$_2$ at valley $K$ with mini Brillouin zone. b) Wannier orbital charge density $|W_{\bm R\ell}(\bm r)|^2$ for the topmost valence bands of the top and bottom layer, respectively, obtained from the continuum model. Black solid line shows the moir\'e unit cell. Calculations are performed using parameters specified in Appendix~\ref{appendix:Wannier_hetero} and setting $(V_d,\psi_d)=(2{\rm meV},-106^\circ)$ and the energy offset to $\Delta E_g=45$meV. }    \label{fig:wannier_continuum_model}
\end{figure}

\section{Lattice Model on the moir\'e scale}
\label{sec:continuum_modeling}

The Wannier orbitals for the two highest energy valence bands of the heterostructure are shown in Fig.~\ref{fig:wannier_continuum_model} and form a honeycomb lattice where the two different sites belong to the upper (MoTe$_2$) and lower (WSe$_2$) layers~\cite{Devakul_2021,Guerci_2023,Crepel_2023,Fengcheng_2023_hetero,PRX_FW_2023,Crepel_2024}; see Appendix~\ref{appendix:Wannier_hetero} for a detailed derivation of the lattice model. 
\begin{equation}\label{tight_binding_hamiltonian}
\begin{split}
    &H = - t_{d}\sum_{\langle \bm r,\bm r'\rangle\in \Lambda_d}\sum_\sigma  c^\dagger_{\bm r\sigma} e^{-i\nu_{\bm r,\bm r'}\frac{2\pi \sigma}{3}} c_{\bm r'\sigma}\\
    &-t_{u} \sum_{\langle \bm r,\bm r'\rangle\in \Lambda_u} f^\dagger_{\bm r} f_{\bm r'}-t_\perp\sum_{\langle\bm r,\bm r'\rangle}f^\dagger_{\bm r}c_{\bm r'}\\
    &- \Delta(N_u-N_d)/2 +U_u\sum_{\bm r\in\Lambda_u} n_{\bm r\uparrow} n_{\bm r\downarrow},
\end{split}
\end{equation}
where $\Lambda_{u/d}$ refers to the upper and lower layer triangular lattices generated by the primitive lattice vectors $\bm a_{1/2}=a_0(\pm\sqrt{3}/2,1/2)$, respectively. For short-handedness, the spin index has been suppressed in terms that are spin independent and Hermitian conjugate is implied.  
Here we have not explicitly written the interaction in the WSe$_2$ layer, which is believed to be smaller than the WSe$_2$ bandwidth, see Appendix~\ref{app:coulomb} for an estimate of the Coulomb integrals, and is therefore not relevant to our considerations here.  

The model has $C_{3z}$ symmetry arising from invariance under three-fold rotations around an axis perpendicular to the planes, time reversal symmetry $\mathcal T=i\sigma^y\mathcal K$, where $\mathcal K$ indicates complex conjugation, U(1)$_v$ symmetry generated by spin $S^z$ rotation and the mirror symmetry $M_{y}$ which sends $\bm r = (x,y) \to(x,-y)$ and $\uparrow\to\downarrow$. $M_{y}$ is an approximate symmetry of the lattice model and is not present in the physical bilayer. 
From Wannierization of the continuum model wavefunctions we find $t_u=4.2$meV and $t_d=8.3$meV while the interlayer hopping is $t_\perp=1.8$meV. The MoTe$_2$ interaction estimated as $U_u=90$meV is the largest energy scale of the problem. At filling $\nu=1$ of the upper layer, the relatively large value $U_u/t_u\approx 20$ means we obtain a Mott insulator of localized magnetic moments in a triangular lattice~\cite{Wu_2018,paul_2024,Morales_2022,Downey_2023,Lee_2023} so that the low energy physics of the MoTe$_2$ layer is governed by magnetic degrees of freedom correlated by a generalized Heisenberg exchange, while the allowed hybridization processes between the MoTe$_2$ and WSe$_2$ layers are charge fluctuations involving creation of a electron in MoTe$_2$ and an hole in WSe$_2$ with double occupation of MoTe$_2$ sites forbidden. The result is that the physics is described by a periodic Anderson model of the form
\begin{equation}
\label{chiral_anderson_model}
    \begin{split}
    H_{\rm PAM}& = \sum_{\bm k}c^\dagger_{\bm k\sigma}\left(\epsilon_{\bm k\sigma}+\frac \Delta 2\right) c_{\bm k\sigma} + J_H \sum_{\langle \bm r,\bm r'\rangle\in \Lambda_u} \bm S_{\bm r}\cdot \bm S_{\bm r'}\\
    &
    -\frac{t_\perp}{\sqrt{N_s}}\sum_{\bm r\in \Lambda_u}\left[\sum_{\bm k} e^{i\bm k\cdot\bm r} V_{\bm k}\mathcal{P}\left(f^\dagger_{\bm r}c_{\bm k}\right)\mathcal{P}+h.c.\right]\\
    &-t_{u} \sum_{\langle \bm r,\bm r'\rangle\in \Lambda_u} \mathcal{P} f^\dagger_{\bm r} f_{\bm r'} \mathcal{P} -\Delta/2\sum_{\bm r\in\Lambda_u} f^\dagger_{\bm r}f_{\bm r},
    \end{split}
\end{equation}
where in intralayer $t_u$ and interlayer $t_\perp$ hopping terms the projector $\mathcal{P}$ suppresses processes that create a double occupied site in the MoTe$_2$ layer. Here the $\bm S$ are spin operators in the MoTe$_2$ layer, $J_H=4t_u t^2_\perp/\Delta^2$ is the superexchange in the charge transfer limit, $N_s$ is the number of unit cells, and $\epsilon_{\bm k\sigma}=-t_d F_{\bm k\sigma}$ with $F_{\bm k\sigma}=2\sum_{j=1}^3\cos\left(\bm \gamma_j\cdot\bm k+2\pi\sigma/3\right)$ and $V_{\bm k}=\sum^{3}_{j=1} e^{i\bm k\cdot{\bm u_j}}$. In the latter expressions, we have introduced the vectors $\bm u_1=a_0(1,0)/\sqrt{3}$ and $\bm u_j=C^{j-1}_{3z}\bm u_1$ connecting the two different triangular lattices and $\bm \gamma_j=C^{j-1}_{3z}\bm a_2$, the primitive vectors of each triangular lattice where $j=1,2,3$.  

\section{Parton representation of the chiral periodic Anderson model} 
\label{sec:chiralPAM}

To account for mixed valence states within the low-energy subspace of zero double occupancy in MoTe$_2$, we employ the slave boson representation $f_{\bm r\sigma}=b^\dagger_{\bm r}\chi_{\bm r\sigma}$~\cite{Barnes_1976,Barnes_1977,read1983solution,Coleman_1984,newns1987mean,Andy1987} with the local constraint $n_b+n_{\chi}=1$. Writing out the effective model~\eqref{chiral_anderson_model} within the parton decomposition leads to
\begin{widetext}
\begin{equation}
\label{chiral_anderson_model}
    \begin{split}
    H_{\rm PAM} =& \sum_{\bm k}c^\dagger_{\bm k\sigma}\left(\xi_{\bm k\sigma}+\Delta/2\right) c_{\bm k\sigma} + J_H \sum_{\langle \bm r,\bm r'\rangle\in \Lambda_u} \bm S_{\bm r}\cdot \bm S_{\bm r'}
    -\frac{t_\perp}{\sqrt{N_s}}\sum_{\bm r\in \Lambda_u}\left[b_{\bm r}\chi^\dagger_{\bm r}\sum_{\bm k} e^{i\bm k\cdot\bm r} V_{\bm k}c_{\bm k}+h.c.\right]\\
    &+ \sum_{\bm r\in \Lambda_u}\chi^\dagger_{\bm r}(\lambda_{\bm r}-\mu-\Delta/2)\chi_{\bm r}+\sum_{\bm r\in\Lambda_u}\lambda_{\bm r}b^\dagger_{\bm r}b_{\bm r}-t_{u} \sum_{\langle \bm r,\bm r'\rangle\in \Lambda_u} \chi^\dagger_{\bm r}b_{\bm r}b^\dagger_{\bm r'} \chi_{\bm r'}, 
    \end{split}
\end{equation}
\end{widetext}
$\lambda_{\bm r}$ is the Lagrange multiplier imposing the constraint $n_b+n_\chi=1$, $\mu$ enforces the total filling to $\nu=1+1$. A finite $n_b$ implies a finite density of holons in MoTe$_2$ and interlayer valence fluctuations.

To decouple the exchange interaction we introduce the bosonic field $Q_{\bm r,\bm r'}$ defined along the bonds of the triangular lattice $\Lambda_u$ that takes into account short-range antiferromagnetic correlations without long-range order. Performing the Hubbard-Stratonovich decoupling of the Heisenberg interaction $J_H \bm S_{\bm r}\cdot\bm S_{\bm r'}$~\cite{Grilli_1990,Sachdev_1990} we obtain the partition function $Z=\int \mathcal D(Q^\dagger,Q;c^\dagger,c;\chi^\dagger,\chi;b^\dagger,b;\lambda)e^{-\int^\beta_0 d\tau[ \mathcal L_1+\mathcal L_2 ]}$,
where the effective action $\mathcal L = \mathcal L_1+\mathcal L_2$, with $\mathcal L_1$ and $\mathcal L_2$ given by
\begin{equation}\label{lagrangian_slave}
\begin{split}
    \mathcal L_1 = &\sum_{\bm r\in \Lambda_u}b^\dagger_{\bm r}(\partial_\tau+\lambda_{\bm r}) b_{\bm r}+\chi^\dagger_{\bm r}\left(\partial_{\tau}+\lambda_{\bm r}-\mu-\frac{\Delta}{2}\right)\chi_{\bm r}\\
    &+\sum_{\bm k\sigma}c^\dagger_{\bm k\sigma}\left(\partial_{\tau}+\xi_{\bm k\sigma}+\frac{\Delta}{2}\right) c_{\bm k\sigma},\\
    \mathcal L_2= &-\frac{t_\perp}{\sqrt{N_s}}\sum_{\bm r}\left[\chi^\dagger_{\bm r}b_{\bm r}\sum_{\bm k} e^{i\bm k\cdot\bm r} V_{\bm k}c_{\bm k}+h.c.\right]\\
    -t_u&\sum_{\langle\bm r,\bm r'\rangle\in\Lambda_u}\chi^\dagger_{\bm r}b_{\bm r}b^\dagger_{\bm r'}\chi_{\bm r'}+\frac{2}{J_H}\sum_{\langle\bm r,\bm r'\rangle\in \Lambda_u}|Q_{\bm r,\bm r'}|^2\\
    &+\sum_{\langle\bm r,\bm r'\rangle\in \Lambda_u}\left[Q_{\bm r,\bm r'} \chi^\dagger_{\bm r'}\chi_{\bm r}+h.c.\right].
\end{split}
\end{equation}
The theory is invariant under the U(1) gauge transformation $\chi_{\bm r}\to e^{i\theta_{\bm r}}\chi_{\bm r}$, $b_{\bm r}\to e^{i\theta_{\bm r}}b_{\bm r}$ and $\lambda_{\bm r}\to\lambda_{\bm r}-i\partial_\tau\theta_{\bm r}$, $Q_{\bm r,\bm r'}\to Q_{\bm r,\bm r'} e^{i(\theta_{\bm r'}-\theta_{\bm r})}$ where $\theta_{\bm r}\in[0,2\pi)$. The latter gauge symmetry results in an emergent gauge field $a_\mu$~\cite{Senthil_2004}, $\theta_{\bm r}=\int^{\bm r}d\bm l\cdot\bm a$, which is approximated to a static background at the mean-field level; fluctuations beyond the mean-field level are discussed in the following when needed. 
We now proceed to solve the model~\eqref{chiral_anderson_model} at the mean-field level to obtain the phase diagram of the system.

\section{Mean-field theory}
\label{sec:meanfieldtheory}

The mean-field solution is obtained by taking the saddle point of the Lagrangian with respect to $b_{\bm r}$, $Q_{\bm r,\bm r'}$ and $\lambda_{\bm r}$. The mean field solution is formally justified in a large-$N$ limit of a SU(N) generalization of the model~\cite{Senthil_2004} but is believed to represent the physics of the relevant phases even at $N=2$. We obtain the set of self-consistent equations: 
\begin{equation}
    \label{mean_field_equations}
    \begin{split}
    &Q_{\bm r,\bm r'} = -J_H\langle\chi^\dagger_{\bm r}\chi_{\bm r'}\rangle/2,\\
    &\lambda_{\bm r}\langle b_{\bm r}\rangle  = \frac{t_\perp}{\sqrt{N_s}}\sum_{\bm k}e^{-i\bm k\cdot\bm r} V^*_{\bm k}\langle c^\dagger_{\bm k}\chi_{\bm r}\rangle \\
    &+ t_u \sum_{\langle \bm r' \rangle_{\bm r}}\langle b_{ r'}\rangle\langle\chi^\dagger_{\bm r'}\chi_{\bm r}\rangle, \\
    &\langle b^\dagger_{\bm r}b_{\bm r}\rangle+\langle\chi^\dagger_{\bm r}\chi_{\bm r}\rangle=1,\\
    &\sum_{\bm r}\langle\chi^\dagger_{\bm r}\chi_{\bm r}\rangle + \langle c^\dagger_{\bm r} c_{\bm r}\rangle=2N_s,
    \end{split}
\end{equation}
where in the first equation the sites $\bm r,\bm r'$ are nearest neighbors and in the second equation, the sum $\sum_{\langle \bm r' \rangle_{\bm r}}$ extends over the nearest neighbor of the site $\bm r$ in the upper layer triangular lattice $\Lambda_u$. In the basis $\Psi=(\chi,c)^T$ the resulting mean-field Hamiltonian  reads: 
\begin{equation}
\label{meanfield_hamiltonian}
    H_{\rm mf}=\sum_{\bm k\sigma}\Psi^\dagger_{\bm k\sigma}\begin{pmatrix}
           \bar \xi_{\bm k}+\lambda -\frac{\Delta}{2} & -t_\perp b_0 V_{\bm k} \\ 
           -t_\perp b_0 V^*_{\bm k} & \xi_{\bm k\sigma}+\Delta/2
    \end{pmatrix} \Psi_{\bm k\sigma},
\end{equation}
where we considered the mean-field solution invariant under translations~\eqref{mean_field_equations}, $\langle b_{\bm r}\rangle=b_0$, $\bar\xi_{\bm k}=- t^*_uF_{\bm k}-\mu$ is the dispersion of the spinons with $F_{\bm k}=2\sum_{j=1}^3\cos(\bm \gamma_j\cdot\bm k)$ and effective hopping amplitude $t^*_u=t_u|b_0|^2-Q$ where we assume the bond variable $Q_{\bm r,\bm r'}$ to be translational $Q_{\bm r,\bm r'}=Q_{\bm r-\bm r'}$ invariant and preserving the point group symmetries of the model. 

The Hamiltonian exhibits two qualitatively different phases. One is a Fermi liquid coexisting with a magnetic state, which may either break translation and spin-rotation invariance, or may be a U(1) spin liquid~\cite{PhysRevLett.90.216403}. In this coexistence phase the local moments are not counted in the Fermi surface volume. Second, there is a heavy fermion phase in which the local moments are hybridized with the itinerant carriers and the Fermi surface includes all of the carriers including the ones that form the local moment.  The $p$-wave nature of the hybridization gives a topological character to the heavy Fermi liquid state leading to a topological spin Hall phase with a spin Hall conductance that is quantized  when a direct gap between the itinerant and the local moment bands is realized~\cite{Guerci_2023}.
\begin{figure}
    \centering
    \includegraphics[width=1\linewidth]{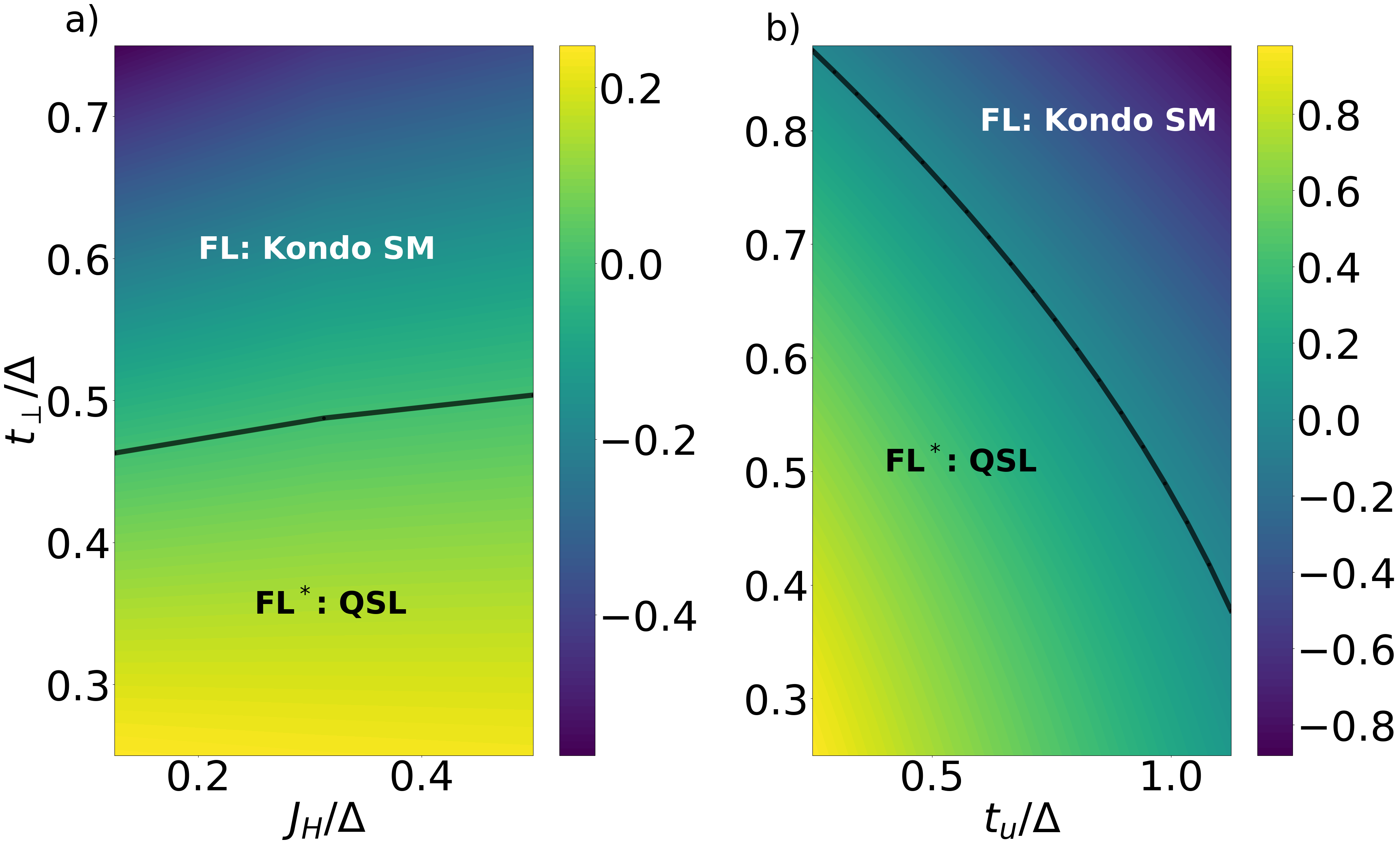}
    \vspace{-0.5cm}
    \caption{Phase diagrams showing competition between the heavy Fermi liquid (FL) and the quantum spin liquid phase (FL$^*$). The color code shows the boson gap $\omega_b/t_d$~\eqref{slave_boson_criticality} becoming gapless at the critical interlayer coupling strength (solid black line) and negative in the Kondo phase. Parameters used are $t_d=8$meV, $\Delta=4$meV and  $k_B T=0.02$meV. Additionally, in a) we set $t_u=4$meV and in b) $J_H=1$meV. }
    \label{fig:phase_diagram_critical_JH_tu}
\end{figure}

In the theoretical approach used here the transition between the two different phases can be determined by finding the critical interlayer tunneling $t_\perp$ where the mass of the holon fluctuations $b_{\bm r}$~\eqref{lagrangian_slave} vanishes. The interlayer charge gap which corresponds to the mass of the holon~\cite{Andy1987,Coleman_1987_prb,Raimondi_1993,Dao_2017,Riegler_2020} is obtained computing Gaussian fluctuations around the mean-field solution $b_0=0$, the Green's function of the holon fluctuations reads: 
\begin{equation}
\label{holon_GF_FLstar}
    G^{-1}_{b}(\bm q,i\omega) = i\omega -\lambda - \omega_{\bm q} -\Sigma_b(\bm q,i\omega),
\end{equation}
where the dispersion relation $\omega_{\bm q}$ reads: 
\begin{equation}
    \omega({\bm q}) = -\frac{t_u}{N_s}\sum_{\bm p} F_{\bm q+\bm p} \langle\chi^\dagger_{\bm p}\chi_{\bm p}\rangle,
\end{equation}
and the self-energy is given by $\Sigma_b(\bm q,i\omega)=t^2_\perp\Pi_{\chi c}(\bm q,i\omega)$. $\Pi_{\chi c}$ is the hybridization susceptibility; setting $p=(\bm p,i\epsilon)$ yields 
\begin{equation}
    \Pi_{\chi c}(q)=\frac{T}{N_s}\sum_{\bm p\, \epsilon}\sum_{\sigma} |V^*_{\bm p}|^2G_{c\sigma}(p)G_{\chi\sigma}(p+q).
\end{equation}
The critical value of the interlayer coupling $t_\perp$ is obtained by the vanishing of the holon mass: 
\begin{equation}
\label{slave_boson_criticality}
   \omega_b(\bm q=0) =\lambda+\omega(\bm q=0)+\Sigma_{b}(\bm q=0,\omega=0)=0.
\end{equation}
In this parton theory the $b_0=0$ phase is of the  FL$^\star$ type of Fermi liquid, which has no broken translational or spin rotation symmetry (so the Mo-layer spins are in a U(1) spin liquid state) but has a Fermi surface of long-lived electronlike quasiparticles whose volume does not count the local moments in the MoTe$_2$ layer. An alternative to this state, not studied here, would involve a Mo-layer in which the spins have a conventional magnetic order. The qualitative dependence of the phase boundary on the model parameters would be similar.

Figs.~\ref{fig:phase_diagram_critical_JH_tu}a) and~\ref{fig:phase_diagram_critical_JH_tu}b) show the critical line (solid black line) where the transition from the U(1) spin liquid to the topological Kondo phase takes place as a function of the microscopic parameters of the model. The phase diagrams are characterized by two different regions: when the holon fluctuations $\omega_b>0$ are gapped the itinerant carriers of the WSe$_2$ layer coexist with a Fermi liquid of spinons (FL$^*$) with short range antiferromagnetic correlations in MoTe$_2$. For $\omega_b<0$, the spin liquid is unstable towards hybridization with the itinerant electron layer developing the heavy Fermi liquid phase (FL). Fig.~\ref{fig:phase_diagram_critical_JH_tu}a) shows that increasing the superexchange $J_H$, which increases the bandwidth of the spinons $\chi$, enhances the critical value of the interlayer hopping to form the heavy Fermi liquid phase. On the other hand, Fig.~\ref{fig:phase_diagram_critical_JH_tu}b) shows the critical line in the plane $t_\perp/\Delta$ and $t_u/\Delta$ with $J_H/\Delta=0.25$. Increasing $t_u$ favors the formation of holes in the local moment layer, thus, inducing the heavy Fermi liquid mixed valence phase where carriers of the local moment layer are promoted to the charge transfer band. 
\begin{figure}
    \centering
    \includegraphics[width=\linewidth]{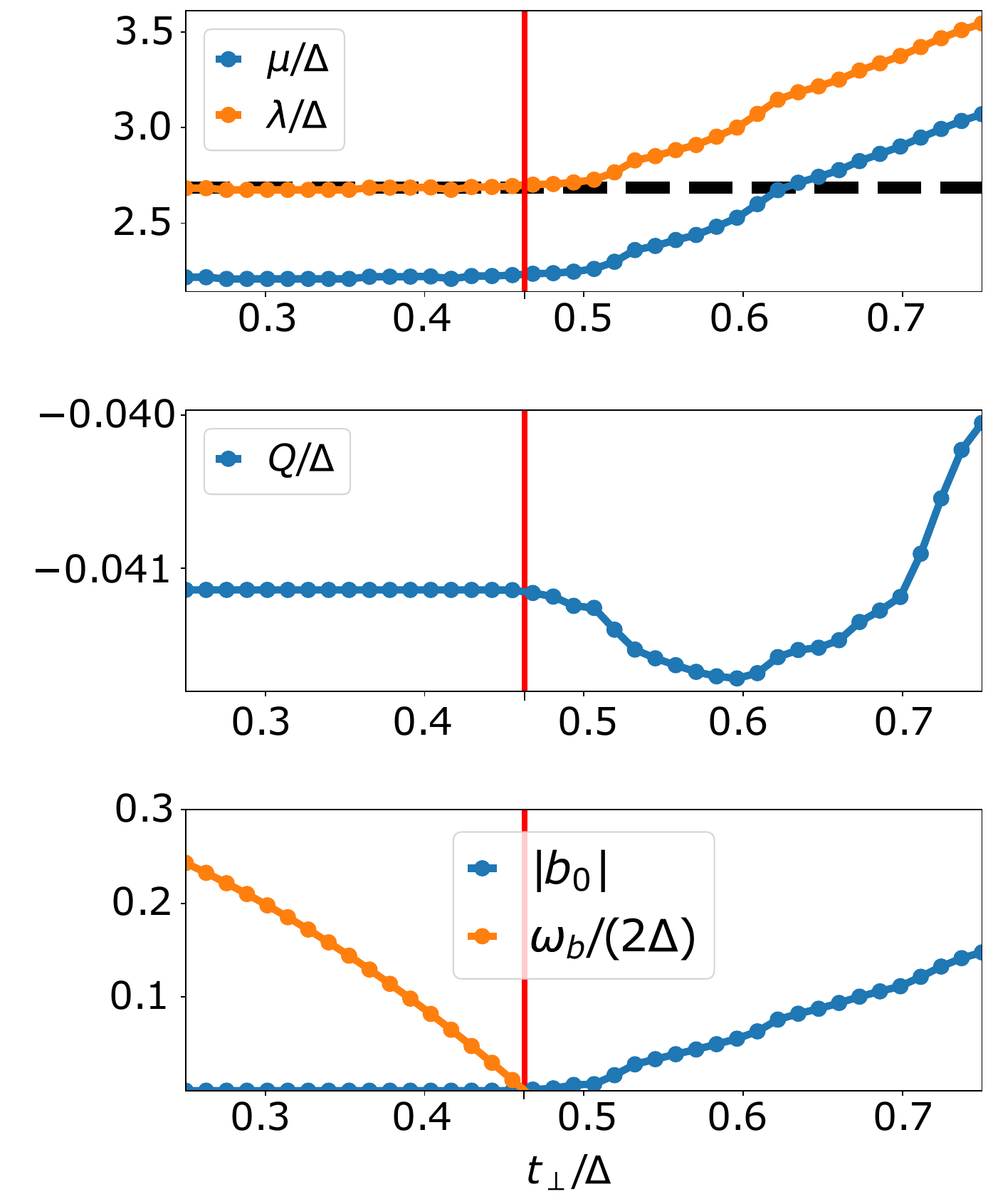}
    \vspace{-0.5cm}
    \caption{From top to bottom we show the evolution of the self-consistent parameters indicated as legends in the panels as a function of $t_\perp/\Delta$. The red vertical line denotes the critical value of $t^c_\perp$ for $J_H/\Delta=0.25$ in Fig.~\ref{fig:phase_diagram_critical_JH_tu}b). In the top panel, the dashed black line showing the value $\mu+\Delta/2$ which well approximates $\lambda$ in the FL$^*$ regime. Parameters of the calculation are $t_u=4$meV, $t_d=8$meV, $k_BT=0.02$meV, $\Delta=4$meV and $J_H=1$meV.}
    \label{fig:line_cut}
\end{figure}

Finally, Fig.~\ref{fig:line_cut} shows the variation of the mean field parameters $b_0$, $Q$, $\mu$ and $\lambda$ as well as the mass of the holon propagator $\omega_b(\bm q=0)/\Delta$~\eqref{slave_boson_criticality} as $t_\perp$ is varied at fixed values of the other parameters. We notice that the formation of the Kondo phase takes place for $t_\perp>t^c_\perp\simeq0.46 \Delta$ with the formation of a condensate of holons $b_0\neq 0$. Correspondingly, the mass of the holon fluctuations vanishes when $\omega_b(\bm q=0)=0$, which is associated with the fluctuations of the phase of the holon degrees of freedom $b_{\bm r}$. Furthermore, in the FL$^*$ regime $\lambda\approx\mu+\Delta/2$, pinning the energy of the spinons $\chi_{\bm r}$ at the Fermi level. Entering into the Kondo regime, $\lambda$ deviates from this value from the non-vanishing hybridization with the itinerant carriers. 
 
\section{Topological Kondo semimetal}
\label{sec:semimetal}

\begin{figure*}[t!]
    \centering
    \includegraphics[width=1\linewidth]{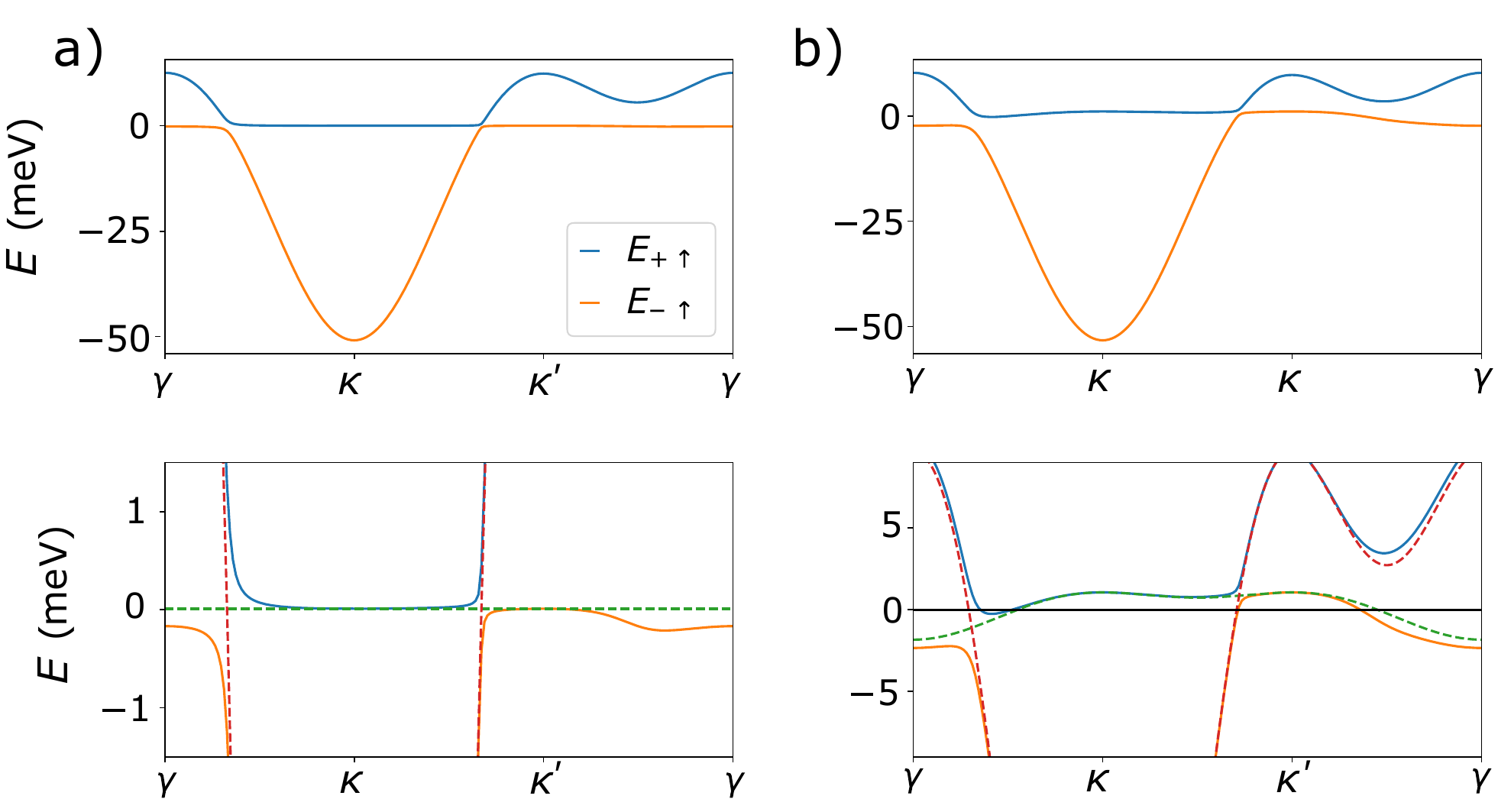}
    \vspace{-0.6cm}
    \caption{ Energy dispersions of the upper and lower quasiparticle bands in the Kondo semimetal phase, plotted along an high symmetry path in the moir\'e Brilloun zone, shown for a wide energy range (upper panels) and near the Fermi energy (lower panels) for $J_H=t_u=0$ (panel a) and $J_H=0.5$meV and $t_u=4$meV (panel b). The indirect gap vanishes in a) and is negative in b) leading to electron hole pockets. This is shown more clearly in the lower panel where we zoom around the Fermi energy and we display with dashed lines the dispersion of the spinons (green) and of the itinerant carriers (red).    
    The self-consistent calculations have been performed at $t_d=7$meV, $k_BT=0.01$meV, $\Delta=4$meV and $t_\perp/\Delta=0.8$.  The kinetic energy gain associated with $t_u$ in delocalizing the holons considerably enhance $b_0$, $|b_0|=0.14$ a) while $|b_0|=0.27$ for panel b).}
    \label{fig:dispersion_relation}
\end{figure*}     
\begin{figure*}
    \centering
    \includegraphics[width=0.8\linewidth]{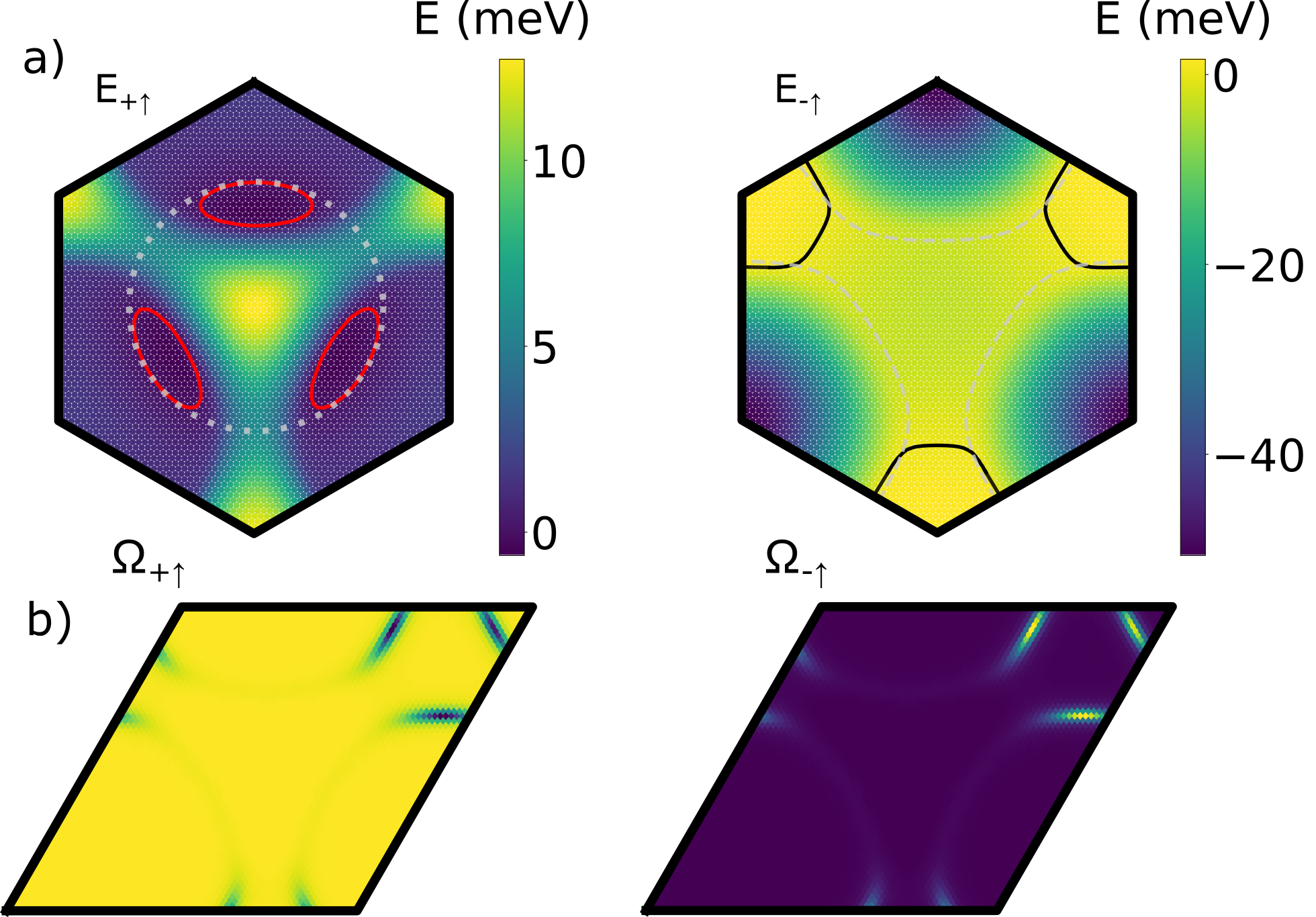}
    \caption{ a) Energy landscape in the first Brillouin zone of the upper (left) $E_{\bm k+\uparrow}$ and lower (right) $E_{\bm k-\uparrow}$ band obtained in the presence of a non-zero interlayer hybridization. The residual electron and hole pockets are highlighted by red (left) and black (right) lines, respectively. To show the original Fermi surface and Fermi surface reconstruction we also display the original Fermi surfaces as a white dashed line. b) Berry curvature distribution in the mini Brillouin zone. Left and right panel refer the upper $E_{\uparrow+}$ and lower quasiparticle band $E_{\uparrow-}$, respectively. The data are obtained for $t_u=4$meV, $t_d=7$meV, $k_BT=0.01$meV, $\Delta=4$meV, $J_H=0.5$meV and $t_\perp/\Delta=0.8$ where $|b_0|=0.27$. 
     }
    \label{fig:2D_map}
\end{figure*}

We now turn our attention to the characterization of the electronic properties of the two different phases. The topological character of the Kondo semimetal phase originates from two ingredients ~\cite{Guerci_2023} directly inherited from the TMD monolayers ~\cite{kormanyos2015k}: the opposite parity (i.e., different $C_{3z}$ eigenvalues of the Wannier orbitals residing in the two layers) and the spin-momentum locking. These two properties lead to a nontrivial momentum dependence of hybridization $V_{\mathbf k}$~\eqref{chiral_anderson_model} which vanishes at the high symmetry points $\kappa$ and $\kappa'$ of the mini Brillouin zone and has a chiral structure. To understand the semimetallic nature of the Kondo phase we start from the simplified limit $t_u=J_H=0$ where the spinon $\chi$ band is exactly flat. The vanishing of $V_{\bm k}$ implies $E_{\kappa\uparrow/\kappa'\downarrow +}=E_{\kappa'\uparrow/\kappa\downarrow-}$. As a result the indirect gap in the quasiparticle band structure $E_{\bm k\sigma\pm}$ vanishes and the state is a nodal semimetal with nodes at $\bm k$ and with a very weak dispersion away from the nodes as shown in Fig.~\ref{fig:dispersion_relation}a). 
Introducing $J_H,t_u$ gives a non vanishing hopping $t^*_u=t_u|b_0|^2-Q>0$ leading to a spinon dispersion with maxima at $\kappa$ and $\kappa'$ where the hybridization function $V_{\bm k}$ vanishes. Thus, the maximum of the lower quasiparticle band $E_{\bm k \uparrow -}$ is located at $\bm k=\kappa$ with value $E_{\kappa'\uparrow-}=3t^*_u+\lambda-\mu-\Delta/2$ and  $E_{\kappa\uparrow+}=E_{\kappa'\uparrow-}$. 
Furthermore, expanding the upper quasiparticle dispersion around $\kappa$ we find $E_{\kappa+\bm k\uparrow +}\simeq E_{\kappa \uparrow +}-\frac{\hbar^2 k^2}{2m_{\rm eff}}$ with $\hbar^2/m_{\rm eff}=3|b_0|^2\left[t_u-t^2_\perp/(6t_d+\lambda-\Delta)\right] +3|Q|$. For realistic model parameters we find $t_u>t^2_\perp/(6t_d+\lambda-\Delta)$ such that $m_{\rm eff}>0$, which implies that the upper and lower quasiparticle bands form electron and hole pockets giving rise to a compensated semimetal. The formation of electron and hole pockets is displayed in Fig.~\ref{fig:dispersion_relation}b) where we show the band structure $J_H,t_u\neq0$. The lower band crosses the Fermi energy around $\kappa'$ forming hole like pockets, while the upper one gives rise to an electron pocket along the $\gamma-\kappa$ line. A two-dimensional visualization of the Fermi surfaces of the compensated heavy Fermi liquid is given in Fig.~\ref{fig:2D_map}a) where the color code shows the dispersion of the upper and lower bands. The electron and hole like residual Fermi surfaces are shown as red and black solid lines in Fig.~\ref{fig:2D_map}a), respectively, while the original Fermi surfaces before hybridization are shown as dashed grey lines. 

In addition to the electron-hole pockets the chiral hybridization gives rise to a Berry curvature flux $\Omega_{\bm k\sigma\pm}=-2\Im\braket{\partial_{k_x}u_{\bm k\sigma\pm}}{\partial_{k_y}u_{\bm k\sigma\pm}}$, where $\ket{u_{\bm k\sigma\pm}}$ are the Bloch vectors of the Hamiltonian~\eqref{meanfield_hamiltonian}, piercing the residual Fermi surfaces of the electron-hole excitations. Notice that due to the time-reversal symmetry $\Omega_{\bm k\uparrow\pm}=-\Omega_{-\bm k\downarrow\pm}$ with spin Chern number $C_{\uparrow\pm}=\mp1$, a non-vanishing density of holons $b_0\neq0$ opens a topological gap in the spectrum. 
Fig.~\ref{fig:2D_map}b) shows the Berry curvature distribution in the mini Brillouin zone. The Berry curvature is concentrated along the direct minimum of the gap contour where the Fermi surface of the itinerant carriers was located. The presence of a residual electron and hole pockets in the Kondo semimetallic state implies that the spin Hall conductivity 
\begin{equation}
\label{spin_Chern_number}
    \sigma^{\uparrow/\downarrow}_{xy} = \frac{e^2}{h}\int_{\rm mBZ}\frac{d^2\bm k}{2\pi}\sum_{n=\pm} \Omega_{\bm k\uparrow/\downarrow n}f(E_{\bm k\uparrow/\downarrow n} )
\end{equation}
is not quantized. For the self-consistent solution shown in Figs.~\ref{fig:2D_map}a) and~\ref{fig:2D_map}b) the spin Hall conductivity is $\sigma^{\uparrow/\downarrow}_{xy}\approx\pm0.7$ (in units of $e^2/h$). Deformations of the moir\'e lattice can remove the electron hole pockets and induce a transition from a Kondo semimetal to a quantum spin Hall Kondo insulator. We detail the transition between these two phases in the next section.

\section{Transition from a Kondo semimetal to a quantum spin Hall Kondo insulator}
\label{sec:strain}

We now discuss the effect of moir\'e lattice distortion corresponding to a random distortion of the moir\'e potential resulting in a relative  shift between the two layers~\cite{Vozmediano_2010}. On the lattice model, the distortion corresponds to a shift of the upper layer $\bm r\in\Lambda_u\to\bm r+\bm \phi$ and results in the variation of the three different bonds of the honeycomb lattice reading $\delta d_j/|\bm u_j|\approx1-\bm u_j\cdot\bm\phi/|\bm u_j|^2$ where the expansion holds for small deformations $\bm\phi$. Fluctuations of the bond length give rise to a variation in the overlap between the Wannier orbitals which reads $t_\perp\to t_{\perp ,j}\approx t_\perp(1+\bm u_j\cdot\bm\phi/|\bm u_j|^2)$~\cite{Franz_2022}. 
The random dislocation $\langle\bm \phi\rangle=0$ preserves on average the point group symmetries of the moir\'e lattice model~\eqref{tight_binding_hamiltonian}. $\phi^a$ are distributed according to $P(\bm \phi)=e^{-|\bm\phi|^2/(2\eta^2)}/(2\pi\eta^2)$ implying $\langle\phi^a\phi^b\rangle_{\rm dis.}=\delta_{ab}\eta^2$ where $\eta$ quantifies the standard deviation of the bond length $|\bm u_j|=a_m/\sqrt{3}$. To leading order in the deformation $\bm \phi$ the average interlayer hybridization reads: 
\begin{equation}
    \langle|V_{\bm k}(\bm \phi)|^2\rangle_{\rm dis.} = \sum^{3}_{jl=1}e^{-i\bm k\cdot(\bm u_j-\bm u_l)}\left(1+\eta\hat{\bm u}_l\cdot\hat{\bm u}_j\right),
\end{equation}
where $\hat{\bm u}_j$ and $\alpha$ are expressed in units of the bond length and $V_{\bm k}(\bm \phi)=\sum^3_{j=1}e^{i\bm k\cdot\bm u_j}t_{\perp,j}/t_\perp$. 
\begin{figure}
    \centering
    \includegraphics[width=0.8\linewidth]{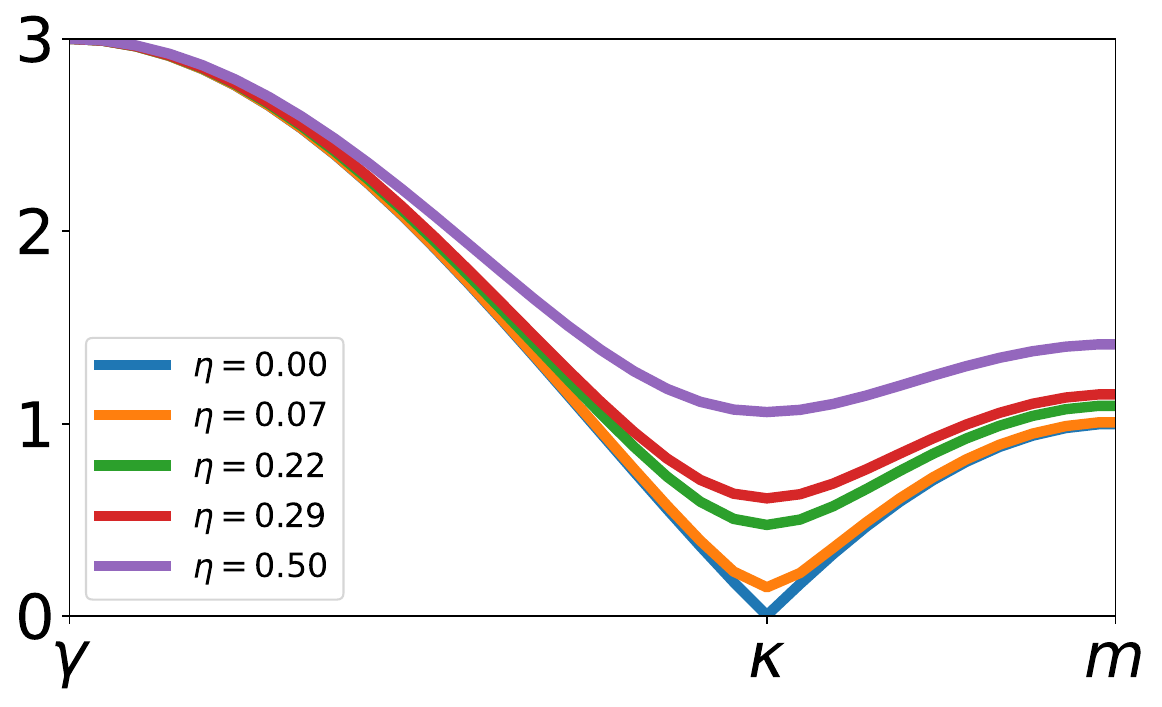}
    \caption{Effect of interlayer disordered deformations on the hybridization function $\sqrt{\langle|V_{\bm k}(\bm \phi)|^2\rangle_{\rm dis.}}$ in the mini Brillouin zone. Different colors correspond to different values of $\eta$ measured in unis of $a_m/\sqrt{3}$, the distance between the two sublattices.}
    \label{fig:Vkdisorder}
\end{figure}
It is instructive to study the effect of interlayer random dislocations on the hybridization function $V_{\bm k}$. Fig.~\ref{fig:Vkdisorder} shows $ \langle|V_{\bm k}(\bm \phi)|^2\rangle_{\rm dis.} $ for different values of $\alpha$. We notice that for $\bm k=\gamma$ the contribution of the random dislocations vanishes since $\sum_{j=1}^3\bm u_j=0$. On the other hand, for $\bm k=\kappa/\kappa'$ we have $\langle|V_{\bm k}(\bm \phi)|^2\rangle_{\rm dis.}=9\eta^2/2$, indicating that random dislocations remove on average the node of the hybridization function. This effect can be readily understood expanding $V_{\bm k}(\bm \phi)$ around $\bm \kappa$  which gives $V_{\kappa+\bm k}(\bm \phi)\approx-3(k_--i\phi_-)/2$ where $k_\pm=k_x\pm ik_y$ and $\phi_{\pm}=\phi_x\pm i\phi_y$ acts as a random gauge field describing a static deformation of the lattice~\cite{Kim_2008,Vozmediano_2010}. For any disorder realization $\bm \phi$, the nodal point of the hybridization function $V_{\bm k}(\bm \phi)$ is shifted by $\bm k_0=(\phi_y,-\phi_x)$. As a result the hybridization pseudogap characterising the chiral Kondo model~\eqref{chiral_anderson_model} is filled by averaging over disorder realizations as shown in Fig.~\ref{fig:Vkdisorder}. We comment that this perturbation is relevant also in the low-doping regime $\nu=1+x$ ($x\ll1$) with $x$ filling factor of the WSe$_2$ layer. This can be readily understood comparing the characteristic Fermi wavevector $k_F=\sqrt{mx/\rho}/\hbar$ with the average deformation introduced by random distortions of the moir\'e lattice which goes like the Kondo hybridization $k_0\propto1/\eta$, the latter being the dominant scale $k_0>k_F$ below a characteristic filling factor $x<x_0\sim\hbar^2\rho k_0^2/m$.
\begin{figure}
    \centering
    \includegraphics[width=1\linewidth]{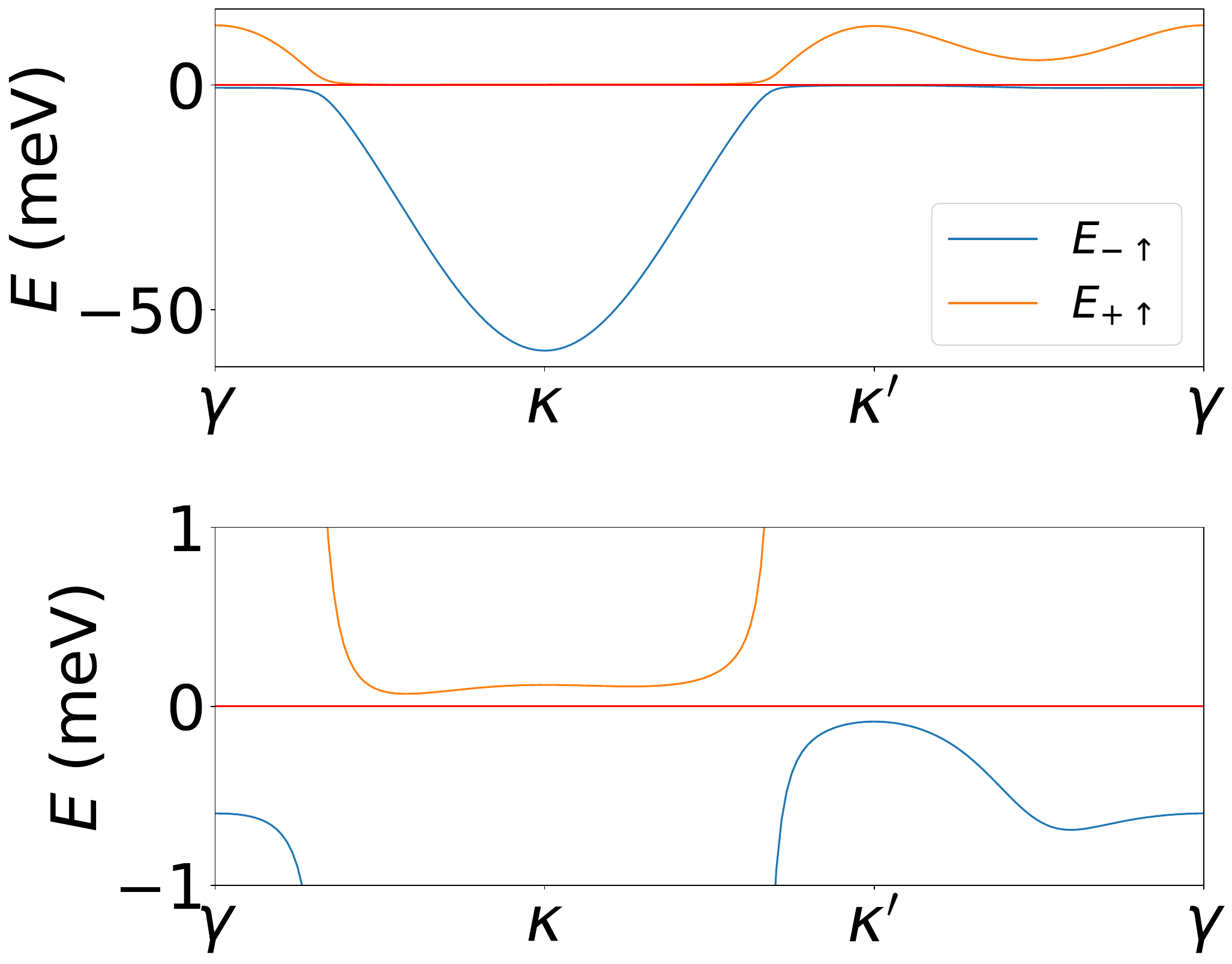}
    \caption{Self-consistent band structure obtained by including the effect of interlayer dislocations for $\eta=0.7$. Top panel shows the dispersion relation of the Kondo phase for spin sector $\uparrow$. Bottom panel shows a zoom of the dispersion around the Fermi energy. The spectrum is characterized by a full gap. The spectrum is obtained for $t_u=1$meV, $t_d=8$meV, $k_BT=0.02$meV, $\Delta=4$meV, $J_H=0.1$meV and $t_\perp/\Delta=0.75$ where $|b_0|=0.2$.}
    \label{fig:dispersion_relation_dislocations}
\end{figure}  

\begin{figure}
    \centering
    \includegraphics[width=0.9\linewidth]{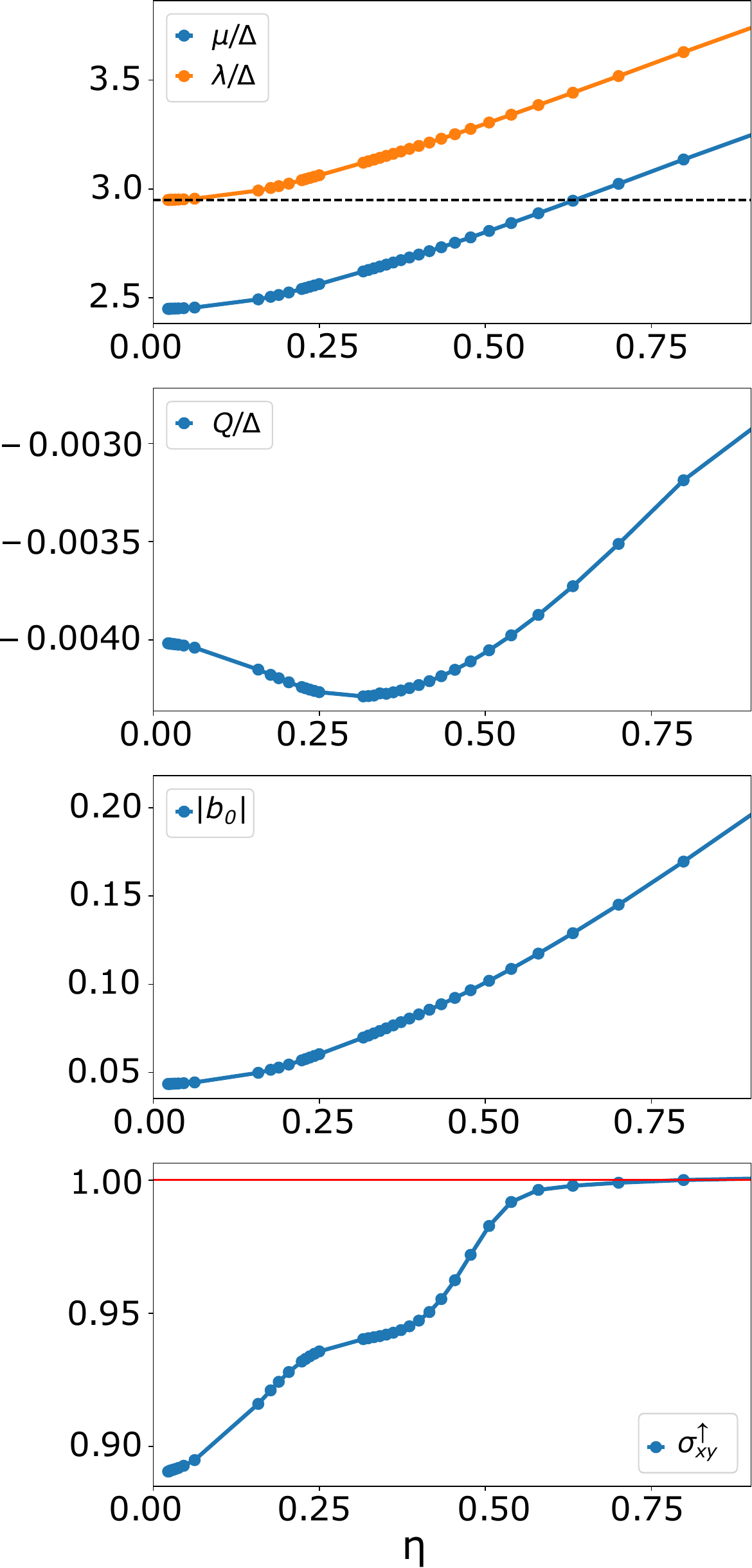}
    \caption{Crossover from a Kondo semimetal to a quantum spin Hall Kondo insulator as a function of the standard deviation of the interlayer bond length $\eta$. Calculations are performed setting $t_u=1$meV, $t_d=8$meV, $k_BT=0.02$meV, $\Delta=4$meV, $J_H=0.1$meV and $t_\perp/\Delta=0.75$. }
    \label{fig:slave_boson_dislocation}
\end{figure}

Fig.~\ref{fig:dispersion_relation_dislocations} shows the band structure in the Kondo regime for large deformation ($\eta=0.7$). Interlayer dislocations give rise to a non-zero gap and removes the electron and hole pockets in the single-particle spectrum. The resulting state is a topological Kondo insulator with quantized spin Hall Chern number $C_{\uparrow/\downarrow}=\pm1$. This can be readily understood in the $J_H=t_u=0$ limit where due to the vanishing of the hybridization $V_{\bm k}$ at $\kappa$ and $\kappa'$ we have $E_{\kappa\uparrow+}=E_{\kappa'\uparrow-}$. Focusing on the spin $\uparrow$ sector, a non-zero distortion $\bm\phi$ gives rise to the $\bm k\cdot\bm p$ Hamiltonian at $\kappa$: 
\begin{equation}
    h_{\uparrow}(\kappa+\bm k)\approx\begin{pmatrix}
        0 & -3it_\perp|b_0|\phi_-/2\\
        3it_\perp|b_0|\phi_+/2 & -\Delta_{\kappa}
    \end{pmatrix} , 
\end{equation}
and at $\kappa'$: 
\begin{equation}
    h_{\uparrow}(\kappa'+\bm k)\approx\begin{pmatrix}
        \Delta_{\kappa'} & 3it_\perp|b_0|\phi_+/2\\
        -3it_\perp|b_0|\phi_-/2 & 0
    \end{pmatrix} , 
\end{equation}
where $\Delta_{\kappa/\kappa'}$ are the direct gaps at $\kappa/\kappa'$, respectively. Applying non-degenerate perturbation theory, we readily find $\delta E_{\kappa\uparrow+}=9\eta^2 t_\perp^2|b_0|^2/(2\Delta_{\kappa})$ and $\delta E_{\kappa'\uparrow-}=-9\eta^2 t_\perp^2|b_0|^2/(2\Delta_{\kappa'})$, which due to level repulsion leads to a full gap in the spectrum:
\begin{equation}
    E_{\rm gap}=E_{\kappa\uparrow+}-E_{\kappa'\uparrow-}=\frac{9\eta^2}{2}\frac{t^2_\perp|b_0|^2(\Delta_\kappa+\Delta_{\kappa'})}{\Delta_\kappa\Delta_{\kappa'}},
\end{equation}
where average over disorder realizations has been performed. The evolution from a topological Kondo semimetal to an insulating state is shown in Fig.~\ref{fig:slave_boson_dislocation} where we study the evolution of the self-consistent solution $(\mu,\lambda,Q,b_0)$ and of the spin Chern number~\eqref{spin_Chern_number} as a function of $\eta$. The bottom panel shows the spin/valley Hall conductance in unit of $e^2/h$ computed with Eq.~\eqref{spin_Chern_number} with $E_{\bm k n\sigma}$ dispersion determined self-consistently. Increasing $\sigma$ increases the interlayer hybridization as shown by the third panel of Fig.~\ref{fig:slave_boson_dislocation}. As a result of the level repulsion, increasing $\eta$ also induces a gap in the spectrum leading to the formation of a quantum spin Hall Kondo insulator with quantized $\Delta\sigma_{xy}=\sigma^\uparrow_{xy}-\sigma^\downarrow_{xy}=2$ in units of $e^2/h$. 

Finally, we conclude summarizing the various phases realized in our theory. Increasing $t_\perp/\Delta$, the system undergoes a transition from a $C_\sigma=0$ FL$^*$ phase to an anomalous spin Hall Kondo phase with $\sigma^\sigma_{xy}\neq 0$. Within, the Kondo regime we find two different states: the semimetal with $\sigma^\sigma_{xy}$ not quantized and the insulator where $\sigma^{\uparrow/\downarrow}_{xy}=\pm 1$ and a small energy gap~\cite{Dzero_2010,Dzero_2016} which is proportional to the holon condensate density $|b_0|^2$ and reflects the energy scale to break the singlets. 
A natural consequence of these non-zero Chern numbers are gapless topological surface states measurable by probing the edge of the sample, which we will study in the next section employing cylindrical geometry.

\section{Edge modes}
\label{sec:edgemodes}

\begin{figure}
    \centering
    \includegraphics[width=\linewidth]{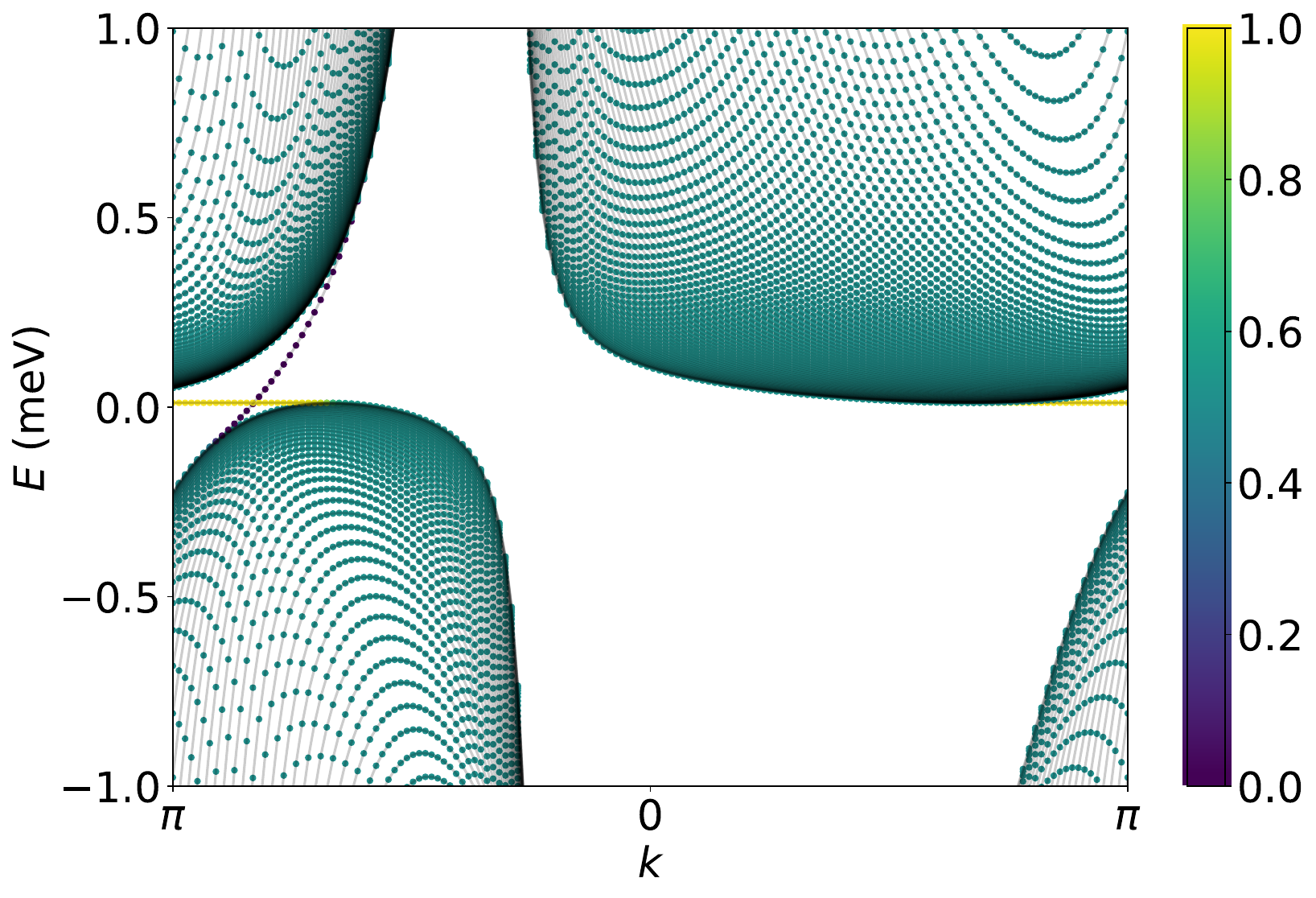}
    \caption{Yellow and blue states are edge modes localized at the boundaries of the cylinder with zigzag boundary conditions for the spin $\uparrow$ sector. The color code shows the localisation of the edge mode at the two boundaries of the cylinder. Calculations are performed setting $t_u=0$meV, $t_d=8$meV, $k_BT=0.02$meV, $\Delta=4$meV, $J_H=0.$meV and to enhance the hybridization amplitude $t_\perp/\Delta=1$. }
    \label{fig:egde_modes}
\end{figure}

We now analyse edge states coexisting with bulk quasiparticles in the Kondo semimetallic regime by solving the mean field equations in a cylinder geometry with periodic boundary conditions along $\bm a_3=\bm a_1+\bm a_2=a_0(0,1)$ and either open in the other direction. The result for open boundary conditions shown in Fig.~\ref{fig:egde_modes} reveals gapless edge modes localised at the two opposite boundaries of the two dimensional cylinder for spin $\uparrow$ carriers; spin $\downarrow$ edge modes are obtained by time reversal symmetry. The two edge modes propagating in opposite directions are characterized by dramatically different velocities: the edge mode localised along the termination of WSe$_2$ has a much larger velocity than the one originating from the local moments of MoTe$_2$. The asymmetry between the two edges originates from the large effective mass of localised carriers in MoTe$_2$ and, generically, from the absence of inversion symmetry which is strongly broken in TMDs. On the single-particle level the edge modes are eigenstates, but many-body scattering processes not considered here can lead to a finite lifetime due to particle-hole exchange with carriers in the semimetallic bands. Also, impurity scattering may couple the edge and bulk modes. A detailed analysis of the response of edge modes to an external electromagnetic field in small gap topological Kondo insulators and semimetals within a self-consistent parton construction approach~\cite{wang2024meanfieldstudyquantumoscillations,wagner2023edgezerosboundaryspinons} is left to future investigations.

\section{Conclusion}
\label{sec:conclusions}

In this work, we have studied a theoretical model of AB-stacked TMD heterobilayers, with parameters chosen to represent MoTe$_2$/WSe$_2$ in the doping regime $\nu =\nu_{Mo}+\nu_W= 1+1$. The relatively strong correlations in the MoTe$_2$ layers and relatively weak correlations in the WSe$_2$ layer naturally lead to two phases: a Kondo-coupled state, and a layer-decoupled state. The principal result of this paper is that the Kondo-coupled state is a Kondo semimetal, with small compensating electron and hole pockets, rather than the Kondo insulator often found at this carrier concentration. Moreover we have shown that the combination  of AB-stacking  and the atomic physics of the constituent materials endows the Kondo semimetal with topological properties arising from a p-wave nature of the interlayer hybridization including edge states that are protected on the single-particle level. Including random interfacial distortions enhances interlayer hybridization and leads to an average gap that produces a topological Kondo insulator. The influence of other sources of disorder, such as Coulomb scattering by charged impurities~\cite{Poduval_2023}, may also be significant, particularly in the regime of low carrier concentration $x\le0.05$ per moir\'e unit cell in WSe$_2$~\cite{zhao2023emergence}.

Models of the general kind considered here may exhibit other phases, including a layer-decoupled phase with a large Fermi surface for carriers in the WSe$_2$ layer and a spin liquid or a magnetic state in the MoTe$_2$ layer. We considered the transition between the Kondo coupled and the FL$^\star$ phase occurring when the MoTe$_2$ system becomes a spin liquid. An alternative is a transition of the familiar `Doniach' kind to a state with magnetic order in the MoTe$_2$ layer. Also as the interaction in the WSe$_2$ layer is increased, a  Mott transition to a fully gapped magnetic insulating state may occur. Investigation of these phases and transitions is an important open question, as is a fuller elucidation of the physics of the edge states. Finally, we highlight that homobilayers, where spin liquid Mott insulators have been theretically proposed~\cite{Zhang_sl_2021,Kiese_2022,kim2024theorycorrelatedinsulatorssuperconductor,Crepel_2024spinon,CSL_Knap_2024}, and trilayer TMDs at finite displacement fields, could also serve as intriguing platforms for the coexistence of itinerant carriers and local moments. In summary, our work shows that  doped and stacked TMD heterobilayers are a novel platform to realize topological Kondo semimetal and insulating states, and  transitions that include a change in topology to other many-body states of current interest. Given the lack of consensus in the heavy fermion community on the topological nature of the putative topological Kondo insulator SmB$_6$~\cite{Alexandrov_2013,Alexandrov_2015,PhysRevB.86.075105,kim2013surface,neupane2013surface,Paglione_2015,nakajima2016one} mostly originating from the experimental observation of 3D bulk quantum oscillations~\cite{li2014two,Tan_2015,baskaran2015majoranafermiseainsulating,PhysRevLett.119.057603,Cooper_2015,varma2024darkfermionsfluctuatingvalence,PhysRevLett.121.026403,Sodemann_2018,Erten_2016,Chowdhury_2018,PhysRevLett.116.046404,blason2024luttingersurfacedominancefermi,wang2024meanfieldstudyquantumoscillations}, having more clear cut platforms with greater theoretical control are of the utmost importance. We expect that a detailed full self-consistent study of the edge modes and their evolution upon doping will help experimentally identify the proposed topological features we have found within.


\begin{acknowledgments}
We thank W. Zhao, T. Senthil, P. Coleman, A. Hardy, A. Georges and N. Wagner for insightful discussions. V.C. thanks K. F. Mak, J. Shan and their entire group for insightful discussions and for their hospitality. 
K.P.L. and J.H.P. are partially supported by NSF Career Grant No.~DMR-1941569. This work was  performed  in part at the Aspen Center for Physics, which is supported by National Science Foundation grant PHY-2210452 (J.H.P.). A.J.M was partially supported by the Columbia University Materials Science and Engineering Research Center (MRSEC), through NSF grant DMR-2011738.
J.C. acknowledges support from the Air Force Office of Scientific Research under Grant No. FA9550-20-1-0260 and the Alfred P. Sloan Foundation.
The Flatiron Institute is a division of the Simons Foundation. 
\end{acknowledgments}

\begin{appendix}

\section{From continuum to lattice model} 
\label{appendix:Wannier_hetero}

In this section we derive the properties of the Wannier orbitals and the lattice model from the continuum moir\'e theory of heterobilayer TMDs. The moir\'e Hamiltonian of TMD heterobilayer at valley $K$~\cite{Wu_2019,Zhang_2021} reads: 
\begin{equation}
\label{H_valley_K}
    H_{K}=\begin{bmatrix}
       -\frac{(\hat{\bm k}-{\bm \kappa})^2}{2m_u}+V_{u}(\bm r) & T(\bm r) \\
        T^*(\bm r) &   -\frac{(\hat{\bm k}-{\bm \kappa}')^2}{2m_d}+V_{d}(\bm r)-\Delta E_g 
    \end{bmatrix},
\end{equation}
where the labels $l=u,d$ refer to top and bottom layers, respectively, and $m_{u/d}$ is the effective mass of the top valence band monolayer. $V_l(\bm r)$ is the intralayer moir\'e potential, $T(\bm r)$ the interlayer tunneling term:
\begin{equation}
    \begin{split}
        &V_l(\bm r)= 2V_l \sum_{j=1}^{3}\cos(\bm g_j\cdot\bm r+\psi_l),\\ 
        &T(\bm r) = t\left(1+\omega e^{-i\bm b_1\cdot\bm r}+\omega^* e^{-i\bm b_2\cdot\bm r}\right),
    \end{split}
\end{equation}
$\omega=e^{2\pi i/3}$ and $\Delta E_g$ is the energy offset. In the latter expression we have introduced 
\begin{equation}
    \bm g_{j}=\frac{4\pi}{\sqrt{3}a_0}\left[\cos \frac {2\pi (j-1)}{3},\sin \frac{2\pi (j-1)}{3}\right]
\end{equation}
with $j=1,2,3$ and $\bm b_{1/2}=4\pi(\pm1/2,\sqrt{3}/2)/{\sqrt{3}a_0}$. Additionally, we set ${\bm \kappa}=\bm q_3$ and ${\bm \kappa}'=-\bm q_2$ where 
\begin{equation}
\bm q_j=\frac{4\pi}{3 a_0} \left[-\sin\frac{2\pi(j-1)}{3},\cos\frac{2\pi(j-1)}{3}\right] 
\end{equation}
with $j=1,2,3$ as depicted in Fig.~\ref{fig:wannier_continuum_model}. We focus on the AB-stacked hetero bilayer composed of MoTe$_2$/WSe$_2$ with model parameters $(a_0,m_u,m_d,t,V_u,\psi_u)=(4.65{\rm nm},0.6m_e,0.35m_e,1.3 { \mathrm{meV}},4.1 { \mathrm{meV}},14^\circ)$ determined by first-principle calculations~\cite{Zhang_2021}. The amplitude of the potential $V_d$ is small~\cite{Zhang_2021}, we set $(V_d,\psi_d)=(2{\rm meV},-106^\circ)$ in our numerical calculations shown in Fig.~\ref{fig:app_continuum_tb}. 

\subsection{Wannier orbitals and hopping amplitudes}
\label{subsec:BW_orbitals}

The Wannier orbitals are obtained by first neglecting the interlayer tunneling potential. This approximation which is justified by the weak interlayer hybridization between the bands. Specifically, we solve the intralayer Hamiltonian 
\begin{equation}
    h_l(\bm r)=-\frac{(\hat{\bm k}-K_l)^2}{2m_l}+V_l(\bm r),
\end{equation} 
with $K_u=\bm \kappa$, $K_d=\bm \kappa'$ and $\bm k=-i\nabla_{\bm r}$ for the topmost Bloch orbital $ \psi_{\bm k l}(\bm r)=e^{i\bm k\cdot\bm r}u_{\bm k l}(\bm r)$ with $u_{\bm k l}(\bm r)$ cell-periodic part. 
The latter can be decomposed in plane waves in the repeated BZ: 
\begin{equation}
u_{\bm k l}(\bm r)=\sum_{\bm G} e^{-i\bm G\cdot\bm r}z_{\bm G l}(\bm k),    
\end{equation}
where $\bm G=n_1\bm b_1+n_2\bm b_2$ and $z_{\bm G l}(\bm k)$ the Fourier amplitudes. Moreover, we set $\bm \kappa=\bm q_3$ and $\bm \kappa'=-\bm q_2$ with $\bm q_1=(\bm b_1+\bm b_2)/3$ and $\bm q_j=C^{j-1}_{3z}\bm q_1$. We observe that the intralayer Hamiltonian $h_l(\bm r)$ transforms under $C_{3z}$ as 
\begin{equation}
h_l(C_{3z}\bm r)=e^{i\bm G_l\cdot\bm r}h_d(\bm r)e^{-i\bm G_l\cdot\bm r} ,
\end{equation}
with $\bm G_l=C_{3z}K_l-K_l$, $\bm G_u=\bm b_2$ and $\bm G_d=\bm b_2-\bm b_1$. As a result, the momentum space projection of the continuum model is invariant under $C_{3z}$:
\begin{equation}
    h_l(C_{3z}\bm k)=D_l(C_{3z})h_l(\bm k)D^\dagger_{l}(C_{3z})
\end{equation}
where $D_{l}(C_{3z})$ is the representation of the $C_{3z}$ symmetry in momentum space for the two different layers:
\begin{equation}
    \begin{split}
    &   D_u(C_{3z})=D(C_{3z}) V^{\bm b_2-\bm b_1},\\ &   D_d(C_{3z})=D(C_{3z}) V^{-\bm b_1}.
    \end{split}
\end{equation}  
We notice that $D(C_{3z})_{\bm G,\bm G'}=\delta_{\bm G,C_{3z}\bm G'}$ and $V^{\bm Q}$ is the sewing matrix which acts as a rigid shift of momentum $\bm Q$ in reciprocal space $V^{\bm Q}_{\bm G,\bm G'}=\delta_{\bm G,\bm G'+\bm Q}$.
The $C_{3z}$ symmetry implies: 
\begin{equation}
\label{app_c3z_bloch_state}
    \begin{split}
        &\psi_{\bm k u}(C_{3z}\bm r)=e^{i(\bm b_2-\bm b_1 )\cdot\bm r}\psi_{C^{-1}_{3z}\bm k u}(\bm r),\\
        &\psi_{\bm k d}(C_{3z}\bm r)=e^{-i\bm b_1\cdot\bm r}\psi_{C^{-1}_{3z}\bm k d}(\bm r).
    \end{split}
\end{equation}
From the Bloch orbitals we build the Wannier functions
\begin{equation}
\label{app_wannier}
     W_{\bm R +\bm r_l l}(\bm r) =\frac{1}{\sqrt{N}}\sum_{\bm k}e^{-i\bm k\cdot\bm R}\tilde \psi_{\bm k l}(\bm r),
\end{equation}
where $n$ is the band index, $\bm R=n_1\bm  a_1+n_2\bm a_2$ defines the moir{\'e} triangular lattice, the sum over $\bm k$ is extended over the first BZ and $N$ is the number unit cells. 
In Eq.~\eqref{app_wannier} $\tilde\psi_{\bm k l}(\bm r)$ we impose $\tilde \psi_{\bm k l}(\bm r_l)\in \mathbb R^+$:
\begin{equation}
    \tilde \psi_{\bm k l}(\bm r)= e^{-i\varphi_{\bm k l}(\bm r_l)}\psi_{\bm k l}(\bm r)
\end{equation}
where $\varphi_{\bm k l}(\bm r_l)=\text{Arg}\,\psi_{\bm k l}(\bm r_l)$ with $\bm r_{u}=0$ and $\bm r_{d}=\bm u_1$. The center of the Wannier orbitals for the two different layers are located in the unit cell at the inequivalent positions $\bm r_u=0$ and $\bm r_d=\bm u_1$, see Fig.~\ref{fig:wannier_continuum_model}b) of the main text, forming a honeycomb lattice. Eqs.~\eqref{app_c3z_bloch_state} imply that the Wannier functions tranforms under $C_{3z}$ as:
\begin{equation}
\label{Wannier_orbitals_symmetries}
    \begin{split}
        W_{\bm R u}(C_{3z}\bm r)&=e^{i(\bm b_2-\bm b_1)\cdot\bm r}W_{C^{-1}_{3z}\bm R u}(\bm r),\\
        W_{\bm R+\bm u_1 d}(C_{3z}\bm r)&=\omega e^{-i\bm b_1\cdot\bm r}W_{C^{-1}_{3z}(\bm R+\bm u_1)d}(\bm r),
    \end{split}
\end{equation}
where we use the relation $\bm a_1=\bm u_1-\bm u_3$, the extra factor in the lower layer Wannier function originates from: 
\begin{equation}
    \varphi_{C_{3z}\bm p d}(\bm u_1)=4\pi/3+\varphi_{\bm p d}(\bm u_3).
\end{equation}

The intralayer hopping is readily obtained: 
\begin{equation}
    \begin{split}
    t^l_{\bm R,\bm R'} &= \mel{W_{\bm R l}}{\hat h_l}{W_{\bm R' l}}\\
    &=\frac{1}{N}\sum_{\bm k}e^{i\bm k\cdot(\bm R-\bm R')}\epsilon_{\bm k l},
    \end{split}
\end{equation}
where $\epsilon_{\bm kl}$ is the energy dispersion of the topmost band of layer $l$. $C_{3z}$ symmetry implies $\epsilon_{C_{3z}\bm k l}=\epsilon_{\bm k l}$ and as a result $t^l_{C_{3z}\bm R,0}=t^l_{\bm R,0}$. 
On the other hand, the leading interlayer tunneling is given by
\begin{equation}
    t^{ud}_{0,\bm R+\bm u_{j+1}}=\mel{W_{\bm 0 u}}{T}{W_{{\bm R+\bm u_1}d}}= \omega^* t^{ud}_{0,C^{-1}_{3z}\bm R+\bm u_j}, 
\end{equation}
particularly for $\bm R=0$ we have $t^{ud}_{0,\bm u_{j+1}}=\omega^*t^{ud}_{\bm u_{j+1}}$. Choosing a gauge where the upper layer is pinned at $\gamma$ and keeping only intralayer and interlayer hopping up to first nearest neighbors we find the tight-binding model in Eq.~\eqref{tight_binding_hamiltonian}. 
\begin{figure}
    \centering
    \includegraphics[width=0.9\linewidth]{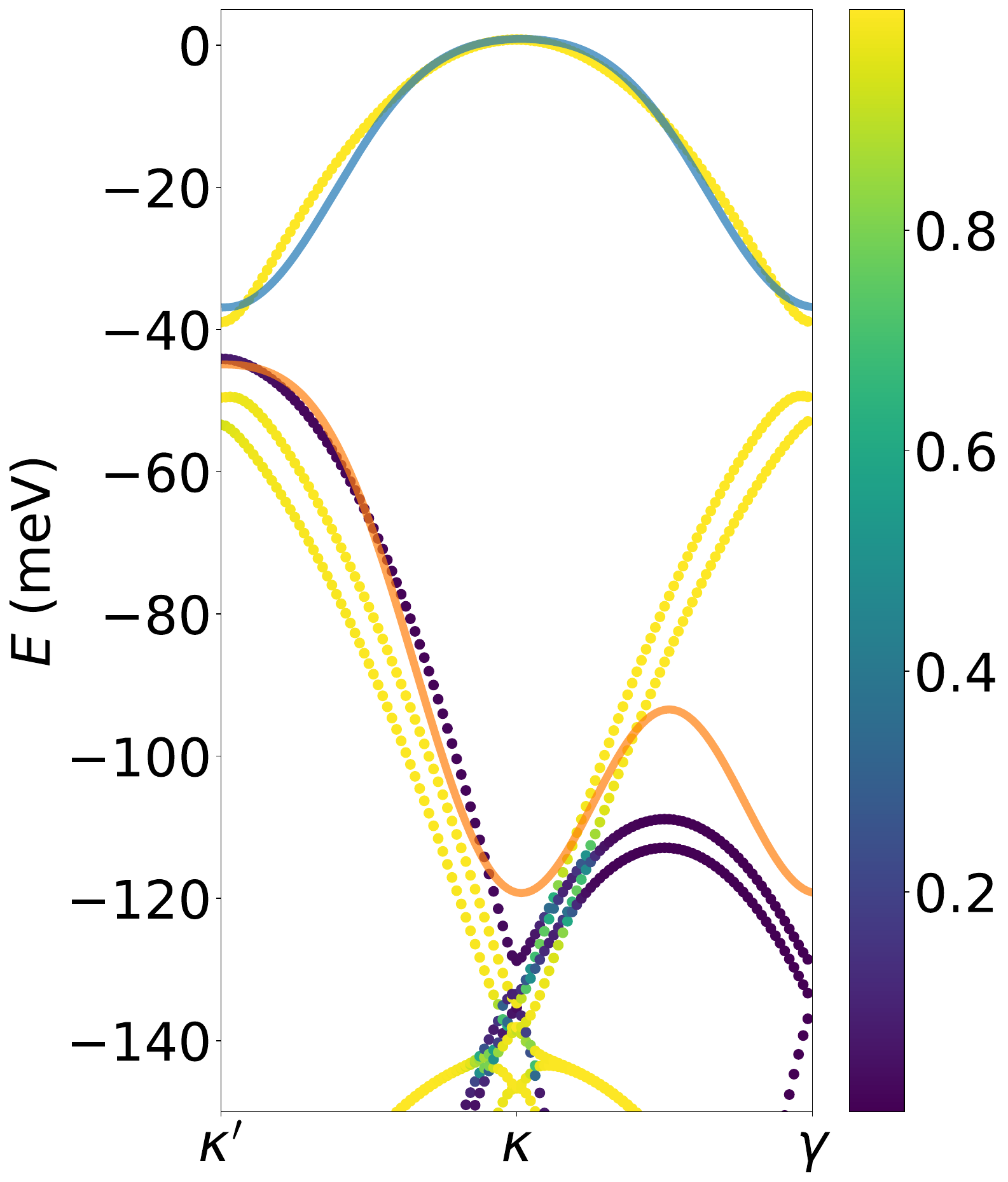}
    \caption{The band structure shows a comparison between the continuum model  and the tight binding Hamiltonian dispersion obtaining by wannierization of the first two bands belonging to opposite layers.}
    \label{fig:app_continuum_tb}
\end{figure}
Figure~\ref{fig:app_continuum_tb} shows a comparison between the band structures of the continuum model and the tight-binding Hamiltonian. Color coding indicates the layer polarization: the upper band is localized in the MoTe$_2$ layer, while the second band from the top is localized in the WSe$_2$ layer. Additionally, we observe the presence of extra MoTe$_2$ bands that partially overlap in energy with the dispersive band of WSe$_2$. In this work, we assume that at a filling factor of 1 in the MoTe$_2$ layer, the intralayer Coulomb repulsion causes these bands to undergo a significant energy shift, larger than the characteristic gap with the WSe$_2$. A detailed analysis of this effect is left for future investigations.

\subsection{Coulomb integrals} \label{app:coulomb}
From the Wannier orbitals $W_{{\bm R}\ell}({\bm r})$ defined in Eq.~\eqref{app_wannier} we compute the strength of the Coulomb interaction. We characterize the interaction through the interaction Hamiltonian projected into the moir\'e basis 
\begin{equation}
    H_{\rm int}=\frac{1}{2}\sum_{ijkl}\sum_{\sigma\sigma'}\langle ij|U|kl\rangle a^\dagger_{i\sigma} a^\dagger_{j\sigma'} a_{l\sigma} a_{k\sigma},
\end{equation}
where: 
\begin{equation}
    \langle ij|U|kl\rangle=\int dx dx'\frac{W^*_i(x)W_k(x)W^*_j(x')W_l(x')}{|x-x'|}.
\end{equation}
For a relative dielectric constant $\epsilon=4.5$ the value of the Coulomb energy scale $e^2/(4\pi \epsilon\epsilon_0 a_0)\approx70$meV. Computing the Coulomb integral of the Wannier orbitals in Eq.~\eqref{app_wannier} we find $U_u=90$meV, $U_d=69$meV and $V=40$meV.

\subsection{Effect of the nearest-neighbor repulsion $V$} 
In this section we comment on the role of nearest neighbor interactions in the regime of interest in our work, which employing 
\begin{equation}
H_{nn} =V\sum_{\langle\bm r,\bm r'\rangle}n_{\bm r} n_{\bm r'}.
\end{equation}
In the $U_u\to \infty$ limit we have utilized the parton decomposition $f_\sigma=b^\dagger\chi_\sigma$ that yields
\begin{equation}
    H_{nn} =V\sum_{\langle\bm r,\bm r'\rangle}(1-b^\dagger_{\bm r}b_{\bm r}) n_{\bm r'},
\end{equation}
naively one could expect that the energy of the holon is reduced by promoting carriers from the localized to the itinerant layer. However, there is also a shift of the energy of the itinerant carriers in the opposite direction. As a result the charging energy will be enhanced by $\mu\to\mu+3V$. The spinons are pinned at the Fermi energy $\lambda\approx\mu+\Delta/2\to\mu+\Delta/2+3V$. Thus, the mass of the boson is left unchanged $\lambda\to \lambda+3V-3V$. From this straightforward estimate we can conclude that $V$ is not relevant in this regime to understand qualitatitve phenomena. However, if we want to determine quantitative estimates of model parameters (e.g. exchange couplings) its important to include nearest neighbor interactions.
\begin{figure*}
\includegraphics[width=1\linewidth]{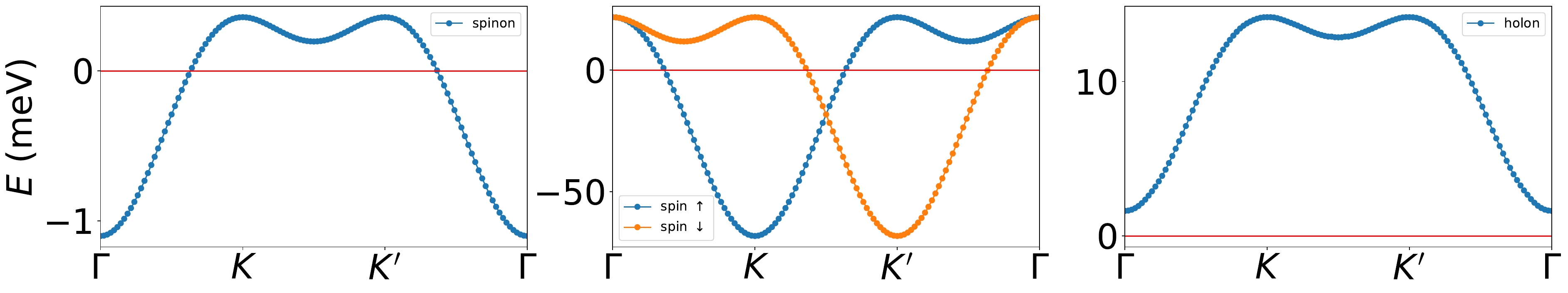}
        \caption{Left and central figures show the dispersion of the spinons, itinerant carriers. Right figure shows the gap of the holon $\omega_b(\bm q)$. The calculations have been performed setting $t_u=4.2$meV, $t_d=5.0$meV, $t_\perp=1.5$meV, $\Delta=3$meV and $J_H=1$meV. The rightmost panel shows the quantity $\Delta+\omega(\bm q)+\Sigma(\bm q,\omega=0)$ which has a minimum at $\bm q=0$ and is characterized by a non-zero gap. The energy gap defines the cost of inducing holon fluctuations. }
        \label{fig:dispersion_excitations}
\end{figure*}


\section{Saddle point equations}
Here we list a series of basic useful results for solving the mean-field problem. At the mean-field level the Hamiltonian~\eqref{meanfield_hamiltonian} leads to the Green's function: 
\begin{equation}
\begin{split}
    \bm G_\sigma(i\epsilon) =&\frac{1}{2}\left[\frac{1}{i\epsilon-d^0_\sigma+|\bm d_\sigma|}+\frac{1}{i\epsilon-d^0_\sigma-|\bm d_\sigma|}\frac{}{}\right]\\
    &+\frac{\bm d_\sigma \cdot\bm\sigma }{(i\epsilon-d_0)^2-|\bm d_\sigma|^2},
\end{split}
\end{equation}
where we have introduced $d^\mu_\sigma(\bm k) = \Tr[\sigma^\mu h^{\sigma}_{\rm mf}(\bm k)]/2$~\eqref{meanfield_hamiltonian}. Notice that time reversal symmetry implies $h^\uparrow_{\rm mf}(\bm k) = h^{\downarrow*}_{\rm mf}(-\bm k)$. Employing the Green's function one can easily compute the single particle density matrix $\langle\Psi^\dagger_a\Psi_b\rangle=G_{ba}(0^{-})$. The hopping of the spinon $Q$~\eqref{mean_field_equations} is given by: 
\begin{equation}
\begin{split}
    Q=\frac{J_H}{48N_s}\sum_{\bm k\sigma s}F_{\bm k}\tanh\frac{\beta(d^0_\sigma+s|\bm d_\sigma|)}{2} \left[1+s\frac{d^z_\sigma}{|\bm d_\sigma|}\right].
\end{split}
\end{equation}
The total filling is constrained to be $2$:
\begin{equation}
    2=\frac{1}{2N_s}\sum_{\bm k\sigma}\left[2-\sum_{s=\pm}\tanh\frac{\beta(d^0_\sigma+s|\bm d_\sigma|)}{2}\right].
\end{equation}
The on-site constraint becomes: 
\begin{equation}
    |b_0|^2-\frac{1}{4N_s}\sum_{\bm k \sigma s }\tanh\frac{\beta(d^0_\sigma+s|\bm d_\sigma|)}{2} \left[1+s\frac{d^z_\sigma}{|\bm d_\sigma|}\right]=0.
\end{equation}
Finally, we have the last saddle-point equation: 
\begin{equation}
\begin{split}
\label{self_consistency_eq}
    b_0=\frac{b_0t^2_\perp/(4N_s)}{\lambda+\frac{12 t_u Q b_0}{J_H}}\sum_{\bm k\sigma s }\frac{|V_{\bm k}|^2}{|\bm d_s|}\tanh\frac{\beta(d^0_\sigma+s|\bm d_\sigma|)}{2}.
\end{split}
\end{equation}
We notice that for $Q<0$ the effective mass $\lambda$ of the holon in Eq.~\eqref{self_consistency_eq} is reduced by the kinetic energy gain $12t_u Q b_0/J_H$.
We conclude providing the expression of the holon dispersion relation in the FL$^*$ phase. Including fluctuations at the gaussian level we find the dispersion \begin{equation}
    \omega(\bm q) = -\frac{t_u}{N_s}\sum_{\bm p\sigma } F_{\bm q+\bm p} f(\bar \xi_{\bm p}+\lambda-\Delta/2),
\end{equation}
and the self-energy: 
\begin{equation}
\Sigma_b(q)=\sum_{\bm p\sigma }\frac{t^2_\perp|V_{\bm p}|^2}{N_s} \frac{f(\xi_{\bm p\sigma}+\frac \Delta 2)-f(\bar\xi_{\bm p+\bm q}+\lambda-\frac \Delta 2)}{i\Omega + \Delta + \xi_{\bm p\sigma} - \lambda - \bar\xi_{\bm p+\bm q}},
\end{equation} 
with $q=(\bm q,i\Omega)$. Fig.~\ref{fig:dispersion_excitations} shows the characteristic behavior of the dispersion of the spinons, itinerant carriers and the holon gap $\omega_b(\bm q)$ in Eq.~\eqref{slave_boson_criticality}.

\end{appendix}

\bibliography{TK_chiral}

\end{document}